\documentclass{aa}
\usepackage{graphicx}
\usepackage{txfonts}
\usepackage{subcaption}
\usepackage{lscape}
\usepackage{placeins}

\begin{document}

   \title{Dynamical plausibility of triple-assisted interaction in Be-star multiple systems}
\author{
C.~A.~Guerrero\inst{1,}\inst{2}\corrauth{carlos.guerrerop@uanl.edu.mx}
\and
O.~V.~Maryeva\inst{3}
\and
M.~Kraus\inst{3}
}
\institute{
Universidad Autónoma de Nuevo León, Facultad de Ciencias Físico Matemáticas,
San Nicolás de los Garza, Nuevo León, México, 66451
\and
Observat\'{o}rio Nacional, Rua General Jos\'{e} Cristino 77, CEP: 20921-400, S\~{a}o Crist\'{o}v\~{a}o, Rio de Janeiro, Brazil
\and
Astronomical Institute, Czech Academy of Sciences,
Fri\v{c}ova 298, 251\,65 Ond\v{r}ejov, Czech Republic
}
   \date{}
\abstract
{Classical Be stars are rapidly rotating B-type stars surrounded by gaseous
circumstellar disks. The origin of the angular momentum required to sustain their
near-critical rotation remains uncertain. Binary interaction is a promising
spin-up channel, and in hierarchical multiple systems a tertiary companion may
drive the secular evolution of an unresolved inner binary.}
{We investigated whether observed Be-star multiple systems are dynamically
compatible with a triple-assisted interaction scenario, in which a resolved
tertiary companion excites the eccentricity of an unresolved inner binary
through the Kozai--Lidov mechanism and brings it closer to an interaction regime.}
{We combined speckle-interferometric detections, reported inner periods, and
proper-motion anomalies (PMas) to construct synthetic populations of
hierarchical triples. Accepted configurations were required to be dynamically
stable, to evolve in the secular regime, to have a Kozai--Lidov timescale
shorter than the adopted main-sequence lifetime scale, and to reach an
inner-periastron interaction proxy. For the best-constrained systems, we
performed refined target-specific simulations using the adopted stellar masses,
distances, periods, PMa scales, and speckle separations.}
{
HD~35079 and HD~55606 provide the cleanest period-plus-speckle dynamical
matches: their observed inner periods and speckle separations are
simultaneously reproduced by stable, secularly active triples.
 HD~205329 remains dynamically plausible, but its
interpretation is astrometrically cautionary because it depends on the PMa
constraint. HD~41335 is a diagnostic case: the period-plus-speckle and
PMa-plus-speckle constraints are viable separately, whereas the simultaneous
period, PMa, and speckle constraints yield no accepted configurations under
the adopted tolerances. 
Several PMa-plus-speckle systems, including HD~37330, HD~41600, HD~205329,
and HD~71510, are observationally promising hierarchical-triple candidates, but
their interpretation is more dependent on the astrometric PMa constraint.
 As an independent consistency check, we estimated the
angular-momentum budget required to spin up the Be star and found a median
fractional requirement of
$\Delta M_{\rm req}/M\simeq4.8\%$ under the adopted assumptions.}
{Several observed Be-star multiple systems admit dynamically plausible
Kozai--Lidov-active configurations in which a resolved tertiary companion can
drive an unresolved inner binary towards interaction. These systems are therefore prime targets for continued
speckle, spectroscopic, and astrometric monitoring, as well as for future
binary-evolution modelling of triple-assisted Be-star spin-up.}
   \keywords{techniques: high angular resolution -- techniques: interferometric -- methods: analytical -- methods: numerical -- binaries: close -- binaries: visual}
   \maketitle
\nolinenumbers

\section{Introduction}
Classical Be stars are rapidly rotating, non-supergiant B-type stars surrounded
by gaseous, dust-free, approximately Keplerian disks
\citep{2003PASP..115.1153P,2013A&ARv..21...69R}. Although rapid rotation is a
defining property of the Be phenomenon, the origin of the angular momentum
required to bring a B-type star close to critical rotation remains an open
problem \citep{2005A&A...440..305F,2013A&ARv..21...69R}. Several channels have been proposed, including single-star evolution, pulsational or magnetic processes, and binary interaction \citep{2016A&A...595A.132Z, 2023Galax..11...83M}. Of these, binary evolution is particularly attractive because mass transfer, tides, and merger events can redistribute angular momentum and produce a rapidly rotating star \citep{2013ApJ...764..166D,2020A&A...641A..42B,2025A&A...698A..38B}.

The binary-interaction picture, however, is not necessarily
limited to isolated binaries. Stellar multiplicity is common,
and the multiplicity fraction increases towards intermediate-
and high-mass primaries \citep{2013ARA&A..51..269D}.
Therefore, a significant fraction of Be stars may have formed
or evolved in hierarchical multiple systems. In such systems,
a distant tertiary companion can perturb an inner binary through
long-term secular dynamics. If the mutual inclination is
sufficiently high, Kozai--Lidov (KL) cycles can excite the
eccentricity of the inner orbit, reducing its periastron distance
and potentially triggering tides, mass transfer, or merger-like
interaction
\citep{1962AJ.....67..591K,1962P&SS....9..719L,2016ARA&A..54..441N}.
The potential role of KL dynamics has been discussed
for the Be binary $\delta$~Sco, where an additional companion
was considered as an explanation for its orbital variability
\citep{2013ApJ...766..119M}.
More broadly, triple-induced secular evolution has been
investigated as a pathway to stellar interactions and mergers
in massive systems, including the proposed triple progenitor
of $\eta$~Carinae
\citep{2016MNRAS.456.3401P,2016ComAC...3....6T}.
Such interactions provide a potential pathway to the spin-up
of Be-star progenitors, although the causal connection between
triple dynamics, a past interaction, present-day rapid rotation,
and the transient appearance of the Be disk remains difficult
to establish observationally.

High-angular-resolution observations offer a way to test whether Be stars occupy architectures compatible with this scenario. In particular, speckle interferometry can resolve companions at subarcsecond separations, which correspond to projected physical separations of tens to hundreds of au for many nearby Be stars \citep[e.g.][]{2020AJ....159..132S,2025AJ....169..251G}. Such companions are generally too wide to be responsible for short-period spectroscopic variability, but they may act as tertiary perturbers of an unresolved inner binary. Conversely, spectroscopic periods and astrometric proper-motion anomalies (PMas) can reveal or suggest inner companions that remain unresolved by speckle imaging. In this context, a PMa refers to the difference between the long-term proper motion measured over the \textsc{Hipparcos}--\textit{Gaia} baseline and the short-term proper motion measured by either \textsc{Hipparcos} or \textit{Gaia}; a significant PMa can be due to a reflex motion induced by an unresolved companion \citep{2019AA...623A..72K}. Combining these observables allows us to ask whether the observed systems are dynamically compatible with secularly active hierarchical triples.

We investigated this question using a synthetic population of hierarchical triples constrained by observations of Be-star multiple systems. We combined three types of information: speckle-detected projected separations, known or suspected inner periods, and PMa information. Our goal was not to reconstruct unique orbital solutions for individual systems. Instead, we ask whether there are stable hierarchical configurations that simultaneously reproduce the available observables and evolve on secular timescales shorter than the adopted main-sequence
lifetime scale, $t_{\rm MS}$, used for the Be star.

The key physical timescale in our analysis is the KL timescale, $t_{\rm KL}$ \citep{1962AJ.....67..591K,1962P&SS....9..719L,2016ARA&A..54..441N}. We define a system as dynamically plausible if it satisfies three conditions: it is hierarchically stable, it has $t_{\rm KL}<t_{\rm MS}$, and the secularly excited inner periastron can approach the stellar-interaction regime. Stability is evaluated using a hierarchical stability criterion appropriate for triple systems \citep{2001MNRAS.321..398M}. In practice, this means that the compatible configurations must allow the inner binary to evolve dynamically within the lifetime of the primary and reach periastron distances comparable to the relevant stellar scale. This criterion does not prove that mass transfer, a merger, or spin-up occurred; it only tests whether the observed architecture is capable of reaching the regime where such processes become possible.

This work focuses on the dynamical step that precedes any detailed
binary-evolution calculation. We ask whether the observed multiple-system
architectures can drive an unresolved inner binary into an interaction regime
within the main-sequence lifetime of the Be primary. Our simulations therefore
establish dynamical plausibility rather than rotational causality. They do not
attempt to compute the subsequent mass transfer, merger evolution, disk
formation, or detailed spin history of the Be star.

After identifying dynamically plausible configurations, we also evaluated
whether the angular-momentum budget required for near-critical rotation is
compatible with modest accreted mass fractions. This additional estimate was
used as a consistency check. Detailed rotational evolution is left to
future binary-evolution calculations.

We applied this framework to a set of Be-star systems with speckle detections
and, for a subset of targets, complementary inner-binary or astrometric
information. The analysis allowed
us to separate systems for which the available observables are compatible with
dynamically active hierarchical triples from cases that are plausible but
limited by astrometric uncertainty or by tension among the adopted constraints.
In particular, the refined analysis highlights HD~35079 and HD~55606 as the
strongest period-plus-speckle dynamical matches, because their independently
constrained inner periods and resolved speckle separations can be reproduced
simultaneously. However, these systems should also be interpreted in light of their individual observational context. 

For HD~35079, the PMa reported by
\citet{2019AA...623A..72K} provides a statistical astrometric
indication of unresolved multiplicity. The companion
masses and orbital scales inferred from such anomalies
depend on the adopted orbital and photocentre
assumptions; consequently, a low-mass or substellar
solution represents a possible interpretation rather
than a uniquely determined companion.
Our period-plus-speckle analysis instead tests the
dynamical compatibility of the independently reported
inner-period and resolved outer-companion scales.

For HD~55606, the inner system is already known as
an interacting Be+sdOB (subdwarf OB) binary \citep{2018ApJ...865...76C}.
Its past mass-transfer history provides the evolutionary
context, while our analysis explores whether the
speckle-resolved companion could have played a
dynamically relevant role as a tertiary perturber.
These systems
therefore serve as period-plus-speckle cases in which the observed inner and
outer scales can be embedded in stable, KL-active hierarchical configurations.

At the same time, systems such as HD~41335, HD~37330,
HD~41600, HD~205329, and HD~71510 remain especially relevant as observational
hierarchical-triple candidates because they combine resolved companions with
independent inner-binary or PMa evidence. HD~41335 emerges as a diagnostic case:
its period-plus-speckle separation and its PMa-plus-speckle separation can each
be modelled independently, but the observed period and the PMa orbital scale
cannot be assigned to the same simple inner orbit under strict tolerances.

The paper is organised as follows. Section~\ref{sec:observational_constraints}
summarizes the observational constraints used in the analysis, including
speckle detections, inner-period information, PMas, and the
target classes adopted for the simulations. Section~\ref{sec:dynamical_framework}
presents the dynamical framework and synthetic population, including the
hierarchical triple model, stability criterion, KL timescale,
interaction proxy, and observational filters. Section~\ref{sec:results}
presents the results of the synthetic matching, including the global secular
timescale behaviour, the consistency with speckle separations, and the
mass-aware refined results for particular systems. Section~\ref{sec:discussion}
discusses the physical interpretation, the special case of HD~41335, the
angular-momentum plausibility estimate, model limitations, and future
observational tests. Section~\ref{sec:conclusions} summarizes the main
conclusions. Supplementary projected-separation distributions and diagnostic
plots are provided in the appendices.

\section{Observational constraints}
\label{sec:observational_constraints}

The purpose of this work is not to derive new astrometric or spectroscopic
measurements, but to test whether the available observables of Be-star
multiple systems are compatible with dynamically active hierarchical
architectures. We therefore used the observational information as a set of
target-specific constraints on synthetic triple configurations.

Only a subset of the systems considered here have independent evidence of
both an inner companion and a resolved outer companion. We therefore use the
term hierarchical triple candidate for systems in which the speckle-resolved
source can be interpreted as a possible tertiary companion to an unresolved
inner binary, rather than assuming that all targets are confirmed triples.

The analysis is organised in two levels. First, we considered the full set of
targets with at least one relevant observable: a speckle-detected companion,
a known or suspected inner period, or a PMa. This
broader sample was used to compare how frequently synthetic hierarchical
systems can satisfy different classes of constraints. Second, we performed
refined target-specific tests for the systems with the strongest combinations
of observables, namely those where a speckle-detected outer companion can be
combined with an inner period and/or a PMa-inferred orbital scale.

\subsection{Speckle-detected companions}
\label{subsec:speckle_constraints}

The relevant observational evidence comes from our two high-angular-resolution surveys of Be stars. The first was carried out at the Pico dos Dias Observatory (OPD/LNA) with the 1.60 m Perkin--Elmer
telescope \citep{2020AJ....159..132S} located in Brazil, and the second corresponds to follow-up
speckle observations obtained with the 2.1 m telescope \citep{2025AJ....169..251G} of the San Pedro Martir Observatory (OAN-SPM) located in Mexico. The PMa information is taken from the
\textsc{Hipparcos}--\textit{Gaia} astrometric analysis of \citet{2019AA...623A..72K}; it is not a
direct-imaging constraint, but an astrometric indication of unresolved
companions through proper-motion differences.
These observations provide the position angle, $\theta$, angular separation, $\rho_{\rm obs}$ in arcseconds between the Be star and the resolved companion, as well as the magnitude difference, $\Delta m$, between the two components. In the context of a hierarchical triple, we interpreted the speckle-resolved
source as a candidate outer companion, $M_3$, orbiting the centre of mass of the
unresolved inner binary, $M_1+M_2$. We denote the semi-major axis of this
relative outer orbit by $a_{\rm out}$, while the unresolved inner binary is
described by an inner semi-major axis $a_{\rm in}$ and inner orbital period
$P_{\rm in}$.

The observed angular separation, however, is not the same as $a_{\rm out}$. Speckle imaging measures an instantaneous projected separation on the plane of the sky:
\begin{equation}
    s_{\perp,{\rm obs}} = \rho_{\rm obs} d,
\end{equation}
where $s_{\perp,{\rm obs}}$ is the projected physical separation and $d$ is the distance to the system in parsecs. With these units, $1''$ at 1 pc corresponds to 1 au, so $s_{\perp,{\rm obs}}$ is obtained in au.

In the synthetic triples, $a_{\rm out}$ is the true semi-major axis of the
relative orbit between the tertiary companion and the centre of mass of the
inner binary. For each synthetic realization, we computed
$s_{\perp,{\rm model}}$ as the instantaneous projected physical separation, on
the plane of the sky, between the synthetic tertiary and the inner-binary
centre of mass. This projected separation depends not only on $a_{\rm out}$,
but also on the outer eccentricity, $e_{\rm out}$, orbital phase, inclination,
argument of periastron, and longitude of the ascending node. We therefore did
not assume that $s_{\perp,{\rm obs}}=a_{\rm out}$. Instead, each synthetic outer
orbit was projected onto the sky and converted to a model angular separation,
$\rho_{\rm model}$, through\begin{equation}
    \rho_{\rm model} =
    \frac{s_{\perp,{\rm model}}}{d}.
\end{equation}
Here, $\rho_{\rm model}$ is the angular separation that the synthetic tertiary
would have in a speckle observation of a system at distance $d$. The distance
($d$) is fixed for each target and corresponds to the adopted \textit{Gaia}-based distance
used in our speckle surveys, taken from \citet{Bailer2021}; it is used
to convert the synthetic projected physical separations,
$s_{\perp,{\rm model}}$, into angular separations, $\rho_{\rm model}$.

\begin{figure}[!t]
    \centering
    \includegraphics[width=\columnwidth]{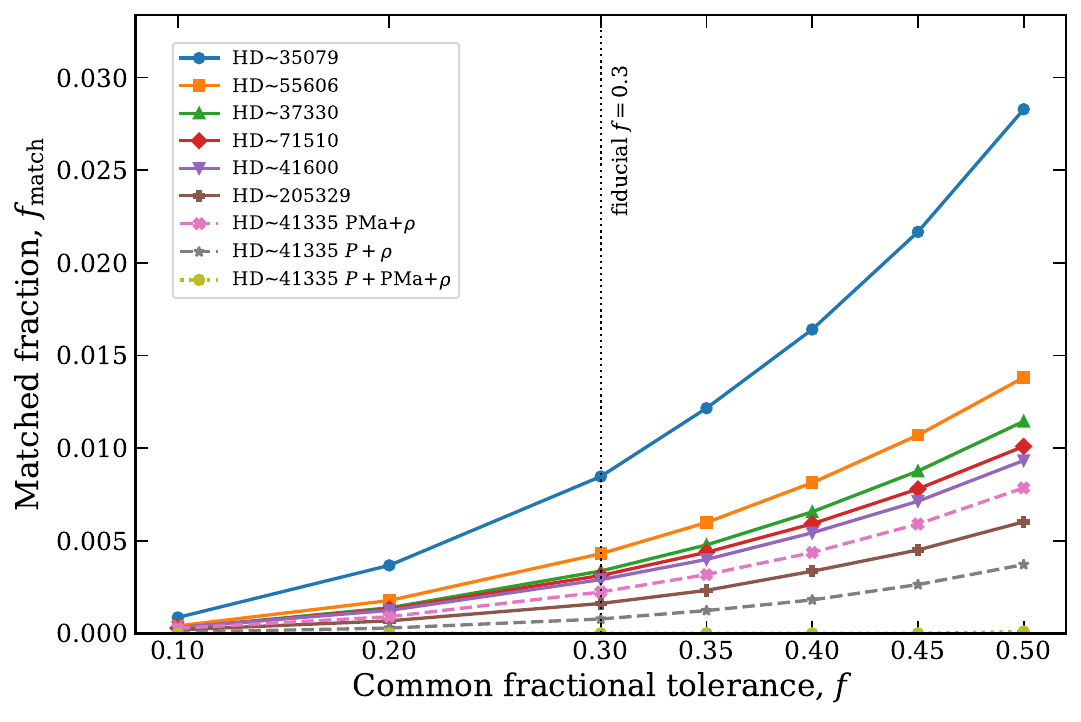}
\caption{
Sensitivity of the refined matched fractions to the
common fractional tolerance ($f$) applied to the available
period, speckle-separation, and PMa constraints.
Each scenario uses the same Monte Carlo population
throughout the tolerance grid. The vertical dotted line
marks the adopted fiducial value, $f=0.3$.
}    \label{fig:tolerance_sensitivity}
\end{figure}

A synthetic configuration is considered compatible with a speckle detection
when $\rho_{\rm model}$ lies within a fractional tolerance around the
observed value:
\begin{equation}
    \rho_{\rm obs}(1-f_{\rho}) <
    \rho_{\rm model} <
    \rho_{\rm obs}(1+f_{\rho}),
\end{equation}
where $f_{\rho}$ is the adopted angular-separation tolerance.
For the refined target-specific simulations, we adopted
$f_{\rho}=0.3$ as a matching tolerance rather than an estimate of
the astrometric measurement uncertainty, which is substantially
smaller. Since orbital phase and projection are explicitly
included in the synthetic separations, this tolerance controls
the allowed deviation between the predicted and observed
angular separations. The same fiducial value, $f=0.3$, is
applied to the available period and PMa constraints to provide
a uniform fractional matching criterion. This choice maintains
a selective correspondence with the observed orbital scales
while accommodating the approximate nature of the PMa-inferred
constraints. Its influence is assessed by varying the common
tolerance over $0.1 \leq f \leq 0.5$, including the intermediate values
$0.35$, $0.40$, and $0.45$, as shown in
Fig.~\ref{fig:tolerance_sensitivity}.

The sensitivity analysis supports $f=0.3$ as a practical
compromise between observational selectivity and the
retention of dynamically viable configurations.
Narrower tolerances substantially reduce the accepted
samples, while broader tolerances admit configurations
with increasingly large deviations from the observed
orbital scales. The main interpretation remains stable
over the tested range: HD~35079 and HD~55606 provide
the most direct period-plus-speckle matches, while
HD~41335 remains a diagnostic case of tension among
the observational constraints.

We also required the model companion to lie within the adopted
speckle-detectability window in angular separation and contrast
(see Appendix~\ref{sec:ap-speckle-detectability}). This window combines the
effective angular-resolution limit, the outer separation limit set by the
field of view, and the approximate contrast sensitivity of the observations.
Thus, a companion is considered accessible to the speckle comparison only when
it falls within the separation and magnitude-difference range probed by the
surveys. This step is important because a non-detection in speckle imaging does not necessarily imply the absence of a companion: the companion may be too close, too faint, or outside the effective instrumental window.

\subsection{Inner periods and spectroscopic constraints}
\label{subsec:period_constraints}

For systems with known or suspected short-period inner companions, we used the reported period as a direct constraint on the inner orbit. These periods are interpreted as tracing an unresolved inner binary rather than the speckle-resolved component. This distinction is essential: a companion detected at tens to hundreds of au cannot generally explain periods of days to months, but it can act as the tertiary perturber of a closer inner binary.

For each synthetic triple, the inner period is computed from the sampled inner
semi-major axis and component masses through Kepler's third law. When an
observed inner period is available, the model is required to satisfy
\begin{equation}
    P_{\rm obs}(1-f_P) <
    P_{\rm in,model} <
    P_{\rm obs}(1+f_P),
    \qquad f_P=0.3
\end{equation}
in the refined analysis. The value $f_P=0.3$ follows the same fiducial
fractional tolerance adopted for the target-specific matching filters, allowing
the period, PMa, and speckle constraints to be applied on a uniform relative
scale. This constraint is applied, for example, to systems such as HD~35079 and
HD~55606, where the combination of a compact inner period and a resolved outer
speckle companion provides a direct test of hierarchical triple compatibility.

\subsection{Proper-motion anomalies}
\label{subsec:pma_constraints}

We also used PMa information from \citet{2019AA...623A..72K}. As stated before, PMa is the difference between the long-term proper motion measured over the \textsc{Hipparcos}--\textit{Gaia} baseline and the short-term proper motion measured by either \textsc{Hipparcos} or \textit{Gaia}. A significant PMa can indicate the reflex motion of the photocentre induced by an unresolved companion.

The PMa information is not equivalent to a full orbit. It provides an
astrometric indication of unresolved multiplicity and an approximate orbital
scale, but its interpretation can be affected by projection, astrometric
systematics, unresolved higher-order multiplicity, and the luminosity ratio of
the components. The luminosity ratio is important because the astrometric motion traces
the photocentre rather than the barycentre; therefore, the same physical orbit
can produce different PMa amplitudes depending on how the light is distributed
between the unresolved components. We denote the PMa-inferred orbital scale by $r_{\rm PMa}$. In the numerical experiment, $r_{\rm PMa}$ is used as an approximate constraint on the inner semi-major axis, $a_{\rm in}$, not as an exact orbital solution.

For systems with a PMa constraint, the synthetic inner semi-major axis is required to satisfy\begin{equation}
    r_{\rm PMa}(1-f_{\rm PMa}) <
    a_{\rm in,model} <
    r_{\rm PMa}(1+f_{\rm PMa}),
\end{equation}
where $f_{\rm PMa}$ is the adopted fractional tolerance. In the refined runs, we also use $f_{\rm PMa}=0.3$. Systems with problematic astrometric indicators, such as high renormalized unit weight error values or implausible PMa-inferred companion masses, are retained only as cautionary or diagnostic cases.

\subsection{Target classes}
\label{subsec:target_classes}

Because the available constraints differ from target to target, we classify the systems into observational classes before applying the dynamical model. This classification is used only to organize the analysis; it does not impose a physical interpretation by itself.

The main classes are:

\begin{itemize}
    \item Period + speckle ($P+\rho$) systems: Targets with a known inner period and a resolved speckle companion. These provide the cleanest test of a hierarchical triple architecture because the inner and outer scales are independently constrained.

    \item PMa + speckle ($\mathrm{PMa}+\rho $) systems: Targets with a resolved speckle companion and a PMa-based indication of an unresolved companion. These systems are dynamically informative but depend on the interpretation of the PMa signal.

    \item Period + PMa + speckle ($P+\mathrm{PMa}+\rho $) systems: Targets for which all three constraints are available. These are highly restrictive and can reveal whether the observed period and PMa scale can plausibly refer to the same inner orbit.

    \item PMa-only systems: Targets without a speckle detection but with significant or marginal PMa evidence. These systems trace possible unresolved companions and are used as a comparison sample for the inner-orbit scale.

    \item Speckle-only ($\rho$-$\mathrm{only}$) systems: Targets with a resolved visual companion but no independent inner-period or PMa constraint. These systems trace the outer projected-separation regime accessible to speckle interferometry and are used as a comparison sample for the tertiary scale.
\end{itemize}

The targets used in the analysis and their associated observational
constraint groups are summarized in Table~\ref{tab:full_target_sample}, while
the systems used for the refined target-specific analysis and their adopted
physical parameters are listed in Table~\ref{tab:refined_targets_params}.
This refined subset includes robust,
cautionary, and diagnostic cases; the systems that drive the main
interpretation are highlighted later in the Results and Discussion. The
stellar masses and radii used as input parameters in the dynamical
calculations are defined in Sect.~\ref{sec:dynamical_framework}, where
published estimates are adopted whenever available.

\begin{table}[ht!]
\caption{Target groups used in the dynamical experiment.}
\label{tab:full_target_sample}
\centering
\begin{tabular}{lcc}
\hline\hline
Group & $N_{\rm obj}$ & Use \\
\hline
$\rho$-$\mathrm{only}$    & 12 & C1 \\
PMa-only                  & 13 & C2 \\
$P+\rho$          & 2  & R1 \\
$\mathrm{PMa}+\rho$             & 4  & R2 \\
$P$ + $\mathrm{PMa}+\rho$    & 2  & R3 \\
\hline
\end{tabular}
\tablefoot{
The table summarizes the observational constraint groups used in the
synthetic triple experiment. Here $N_{\rm obj}$ is the number of observed
targets in each group. Column `Use' indicates the role of each group:
C1 = comparison sample for outer projected separations; C2 = comparison sample
for unresolved-companion signatures; R1 = refined test with independently
constrained inner period and speckle-detected outer companion; R2 = refined
or cautionary test combining a PMa-inferred inner scale with a speckle-detected
outer companion; R3 = diagnostic over-constrained case combining inner period,
PMa, and speckle separation. The
object-by-object list of the refined systems is given in
Table~\ref{tab:refined_targets_params}.
}
\end{table}

The object-by-object inventory of the 33 unique targets, together with the
observational information used for their classification, is provided in
Appendix~\ref{sec:ap-observational-sample}. The counts in
Table~\ref{tab:full_target_sample} refer to unique targets; systems with
multiple resolved visual pairs may contribute more than one speckle separation
to the separation-distribution figures.

\begin{table*}[ht!]
\small
\caption{Refined target-specific systems and adopted physical parameters.}
\label{tab:refined_targets_params}
\centering
\begin{tabular}{lcccccccl}
\hline\hline
Target & $\rho_{\rm obs}$ & $d$ & $P_{\rm obs}$ & PMa &
$M_1$ & $M_2$ & Mass source & Constraint class \\
       & (arcsec) & (pc) & (days) &  &
($M_\odot$) & ($M_\odot$) & & \\
\hline
HD~35079  & 0.09 & 331  & 9.807 & no/marginal
          & $4.918$ & $qM_1,\ q=0.42$--$0.62$
          & S3; S1,S2
          & $P+\rho$ \\

HD~55606  & 0.09 & 957  & 93.8  & no
          & $6.2$ & $0.9$
          & S4
          & $P+\rho$ \\

HD~37330  & 0.29 & 369  & --    & marginal/caution
          & $4$--$7$ & $0.072$
          & S3
          & $\mathrm{PMa}+\rho$ \\

HD~41600  & 0.08 & 349  & --    & marginal/caution
          & $5.176$ & $0.246$
          & S3
          & $\mathrm{PMa}+\rho$ \\

HD~71510  & 0.27 & 213  & --    & marginal
          & $3$--$7$ & $0.029$
          & S3
          & $\mathrm{PMa}+\rho$ \\

HD~205329 & 0.11 & 3435 & --    & yes, caution
          & $8$--$15$ & $0.90$
          & S3
          & $\mathrm{PMa}+\rho$ \\

HD~41335  & 0.53 & 507  & 80.9  & yes
          & $17.6$ & $1.03$
          & S5
          & $P$ + $\mathrm{PMa}+\rho$ \\

HD~76838  & 2.82 & 336  & 3.85  & caution
          & $5$--$10$ & $34.8$
          & S3
          & $P$ + $\mathrm{PMa}+\rho$ \\
\hline
\end{tabular}
\tablefoot{
Systems used in the refined target-specific analysis,
i.e. objects with a speckle-detected companion plus an inner-period
constraint, a PMa constraint, or both. PMa indicates whether a proper-motion anomaly is used
as an additional constraint. Masses are adopted from direct literature
estimates when available and otherwise from broad ranges consistent with the
reported spectral type or PMa-based estimates. Source codes:
S1 = \citet{1983RMxAA...8..115M};
S2 = \citet{1991IzEhn..56...19M};
S3 = \citet{2019AA...623A..72K};
S4 = \citet{2018ApJ...865...76C};
S5 = \citet{2024ApJ...962...70K}.
For HD~35079, $M_1$ is taken from the PMa-based estimate of
\citet{2019AA...623A..72K}, while the adopted mass-ratio interval combines
the spectroscopic estimates of \citet{1983RMxAA...8..115M} and
\citet{1991IzEhn..56...19M}. Stellar radii are used only to normalise the
interaction proxy, $r_{\rm peri,in}/(R_1+R_2)$, and are assigned through an
approximate main-sequence mass--radius scaling, except for the compact sdO
companion in HD~41335. Systems with problematic or marginal astrometric
indicators are retained as cautionary or diagnostic cases rather than as secure
triple detections. HD~76838 is retained as a cautionary diagnostic case because
the PMa-based companion mass is unusually large for the adopted Be-star mass
range.
}
\end{table*}

The systems listed in Tables~\ref{tab:full_target_sample} and
\ref{tab:refined_targets_params} define a hierarchy of observational constraints
rather than a homogeneous set of confirmed hierarchical Be triples. The
simulations below are therefore used to test whether the observed separations,
periods, and/or PMa scales are compatible with dynamically stable hierarchical
architectures capable of producing KL cycles within the adopted
main-sequence lifetime scale. The resulting fractions are interpreted as
relative plausibility measures within the synthetic population, not as
occurrence rates.

\section{Dynamical framework and synthetic population}
\label{sec:dynamical_framework}

The goal of this work is to test whether the observed
architectures are compatible with a dynamically plausible path-
way towards inner-binary interaction rather than to reconstruct a unique evolutionary history for each system. 
We therefore constructed a synthetic population of hierarchical triples and compared it with
the available observables for the Be-star systems in our sample. In this
framework, the inner orbit is defined by the semi-major axis $a_{\rm in}$,
eccentricity $e_{\rm in}$, and period $P_{\rm in}$ of the close pair, while the
outer orbit is defined by the semi-major axis $a_{\rm out}$, eccentricity
$e_{\rm out}$, and mutual inclination $i_{\rm mut}$ of the tertiary companion
with respect to the inner binary, expressed in radians; we treated $i_{\rm mut}$ as one of the orbital
parameters that determines whether a hierarchical configuration can enter the
KL-active regime. This allowed us to identify the subset of architectures that
are compatible with the observed constraints and capable of driving secular
eccentricity excitation. The total mass of the inner binary is
$M_{\rm in}=M_1+M_2$, where $M_1$ denotes the Be star in the currently observed
configuration and $M_2$ denotes the unresolved inner companion. The mass of the
tertiary companion is denoted by $M_3$. Throughout this work, `primary'
refers to the component labelled $M_1$ in the observed system, without
implying that it was the initially more massive star before any past mass
transfer or merger-like interaction.

Whenever available, published mass estimates were adopted for the inner
binary components. In particular, for the systems with inner-period
constraints we used the values reported in the literature and summarized by
\citet{2025AJ....169..251G}: HD~35079, HD~41335, and HD~55606. For systems
without direct mass estimates, we adopted masses consistent with the reported
spectral type and propagated the uncertainty through the synthetic
population. The adopted masses affect the conversion between $a_{\rm in}$ and
$P_{\rm in}$ through Kepler's law, the KL timescale, the
main-sequence lifetime estimate, and the normalization of the interaction
proxy through the stellar radii.

The adopted stellar masses and observational parameters
are summarized in Table~\ref{tab:refined_targets_params}.
The masses enter the analysis through the conversion
between $a_{\rm in}$ and $P_{\rm in}$ using Kepler's law,
and through the KL timescale in
Eq.~(\ref{eq:tkl}). The stellar radii, which are not
tabulated, are estimated from the adopted masses using
an approximate main-sequence mass--radius relation,
with the exception of the compact sdO (subdwarf O) companion
in HD~41335. These radii are used to normalise the
interaction proxy, $r_{\rm peri,in}/(R_1+R_2)$, where $R_1$ and $R_2$ are the stellar radii of the 
primary and inner companion. Thus, the table documents the present-day physical
parameters used to connect the synthetic architectures with the observed
systems.

For each synthetic system, we first required a hierarchical configuration,
$a_{\rm out}>a_{\rm in}$, and then imposed long-term dynamical stability using
the Mardling--Aarseth criterion. Following \citet{2001MNRAS.321..398M}, this
criterion provides a critical ratio between the outer periastron distance and
the inner semi-major axis as a function of the outer mass ratio, outer
eccentricity, and mutual inclination. In the notation adopted here, the outer periastron distance
is\begin{equation}
    r_{\rm peri,out}=a_{\rm out}(1-e_{\rm out}),
\end{equation}
and the stability condition is written as
\begin{equation}
    \frac{r_{\rm peri,out}}{a_{\rm in}}
    \gtrsim
    2.8
    \left[
    \left(1+\frac{M_3}{M_{\rm in}}\right)
    \frac{1+e_{\rm out}}{\sqrt{1-e_{\rm out}}}
    \right]^{2/5}
    \left(1-\frac{0.3\,i_{\rm mut}}{\pi}\right).
    \label{eq:mardling_aarseth}
\end{equation}

\noindent This criterion rejects systems in
which the tertiary approaches the inner binary too closely for the assumed
hierarchical secular description to remain valid.

The main dynamical mechanism considered here is the KL effect
\citep{1962AJ.....67..591K,1962P&SS....9..719L}. At quadrupole order, a
mutually inclined tertiary can drive coupled oscillations of the inner
eccentricity and mutual inclination. The characteristic KL timescale is
estimated as
\begin{equation}
    t_{\rm KL}
    =
    \frac{M_1+M_2+M_3}{M_3}
    \frac{P_{\rm out}^2}{P_{\rm in}}
    \left(1-e_{\rm out}^2\right)^{3/2},
    \label{eq:tkl}
\end{equation}
where $P_{\rm out}$ is the outer orbital period. We used this expression as an order-of-magnitude secular clock and
compared it with the main-sequence lifetime of the Be star, $t_{\rm MS}$,
estimated from the adopted primary mass using a standard analytic
mass-lifetime scaling, consistent with rapid stellar-evolution prescriptions
\citep{2000MNRAS.315..543H}. 

In this work, $t_{\rm MS}$ is used as an order-of-magnitude evolutionary clock
for the dynamical comparison and is estimated from the mass of the present-day
Be component for consistency across the synthetic population. In a full
binary-evolution reconstruction, the relevant clock for the onset of mass
transfer may instead be the lifetime of an initially more massive donor.
Accordingly, the condition $t_{\rm KL}<t_{\rm MS}$ is interpreted here as a
dynamical-timescale plausibility requirement, rather than as a complete
evolutionary age constraint.

A system is considered dynamically relevant for the present problem when\begin{equation}
    t_{\rm KL}<t_{\rm MS}.
    \label{eq:tkl_condition}
\end{equation}
Equivalently, the quantity
\begin{equation}
    N_{\rm KL}=\frac{t_{\rm MS}}{t_{\rm KL}},
\end{equation}
estimates the number of KL timescales available during the main-sequence
lifetime of the primary. A supplementary diagnostic showing how $t_{\rm KL}/t_{\rm MS}$ varies across
the hierarchy ratio $a_{\rm out}/a_{\rm in}$ and the outer eccentricity
($e_{\rm out}$) is presented in
Appendix~\ref{sec:ap-hierarchy-timescale}.

The maximum eccentricity reachable in the quadrupole approximation depends
primarily on the mutual inclination. For initially nearly circular inner
orbits, it can be approximated as\begin{equation}
    e_{\rm max}
    \simeq
    \left[
    1-\frac{5}{3}\cos^2 i_{\rm mut}
    \right]^{1/2},
    \label{eq:emax}
\end{equation}
which is real only inside the classical KL inclination window,
approximately $39.2^\circ \lesssim i_{\rm mut}\lesssim 140.8^\circ$. We used
this expression as a proxy for whether the tertiary can excite the inner binary
to high eccentricity. The corresponding inner periastron distance is\begin{equation}
    r_{\rm peri,in}=a_{\rm in}(1-e_{\rm max}).
    \label{eq:inner_periastron}
\end{equation}

We then defined an interaction proxy by requiring this periastron distance to be
comparable to the physical size of the inner binary,
\begin{equation}
    r_{\rm peri,in}
    \leq
    f_{\rm int}\,(R_1+R_2),
    \label{eq:interaction_proxy}
\end{equation}
where $f_{\rm int}$ is a dimensionless tolerance factor. In the fiducial model we
adopted $f_{\rm int}=3$; additional values were used only as sensitivity
tests. This condition is a dynamical filter identifying systems for which KL
eccentricity excitation can plausibly bring the inner binary into an interaction
regime.

\begin{figure}[!t]
    \centering
    \includegraphics[width=\columnwidth]{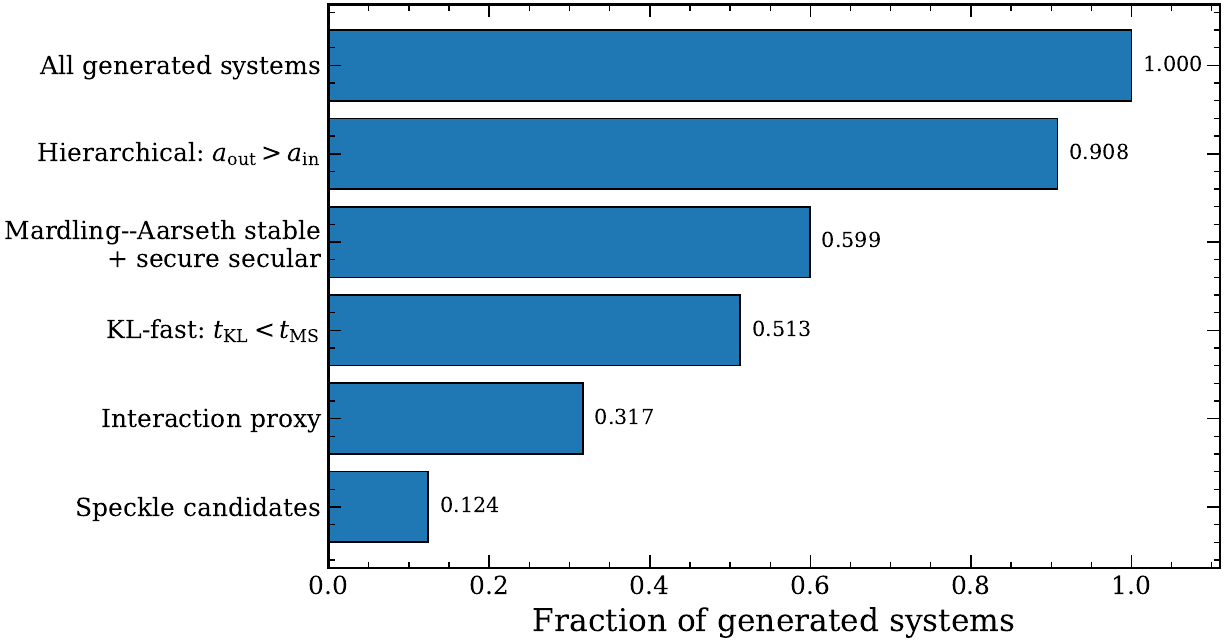}
    \caption{
    Sequential filtering of the synthetic hierarchical-triple population. The
    bars show the fraction of systems surviving each successive dynamical or
    observational requirement, ending with the speckle-detectable channel. The Mardling--Aarseth stability and secure-secular requirements are shown
together because, for the fiducial population, the secure-secular criterion does
not remove additional systems after the stability filter.
    }
    \label{fig:synthetic_cutflow}
\end{figure}

Figure~\ref{fig:synthetic_cutflow} summarizes the sequential effect of these
filters on the synthetic population. The purpose of this figure is to document
the model-selection procedure, not to assign probabilities to individual
targets. Starting from all generated systems, we first imposed hierarchy and
long-term stability. We then required a KL timescale shorter than
the adopted main-sequence lifetime scale. The subsequent filters identify the
systems that satisfy the interaction proxy and remain compatible with the
speckle-detectable channel.

To connect the synthetic systems with the observations, we applied the
object-specific filters defined in Sect.~\ref{sec:observational_constraints}.
For each target, the accepted synthetic configurations must satisfy the
available combination of constraints: the inner-period window when
$P_{\rm obs}$ is available, the projected-separation window when a
speckle-resolved companion is present, and the PMa-inferred inner-scale window
when a PMa is used. 

The outcome of this procedure is an accepted subset of synthetic configurations
for each observational class or target. In the following section, these subsets
are used to quantify the efficiency of the secular channel and to evaluate
which observed systems provide the strongest tests of hierarchical triple
compatibility.

\section{Results}
\label{sec:results}

In this section we apply the dynamical and observational filters described in Sects.~\ref{sec:observational_constraints} and \ref{sec:dynamical_framework}. We first examine the global properties of the accepted synthetic configurations and then evaluate how the observed Be-star systems populate the allowed dynamical channel. Because the matched samples are produced by applying object-specific filters to a synthetic population, they should be interpreted as plausibility regions rather than unique orbital solutions.

\subsection{Global secular timescales}
\label{subsec:global_timescales}

\begin{figure}[!t]
    \centering
    \includegraphics[width=\columnwidth]{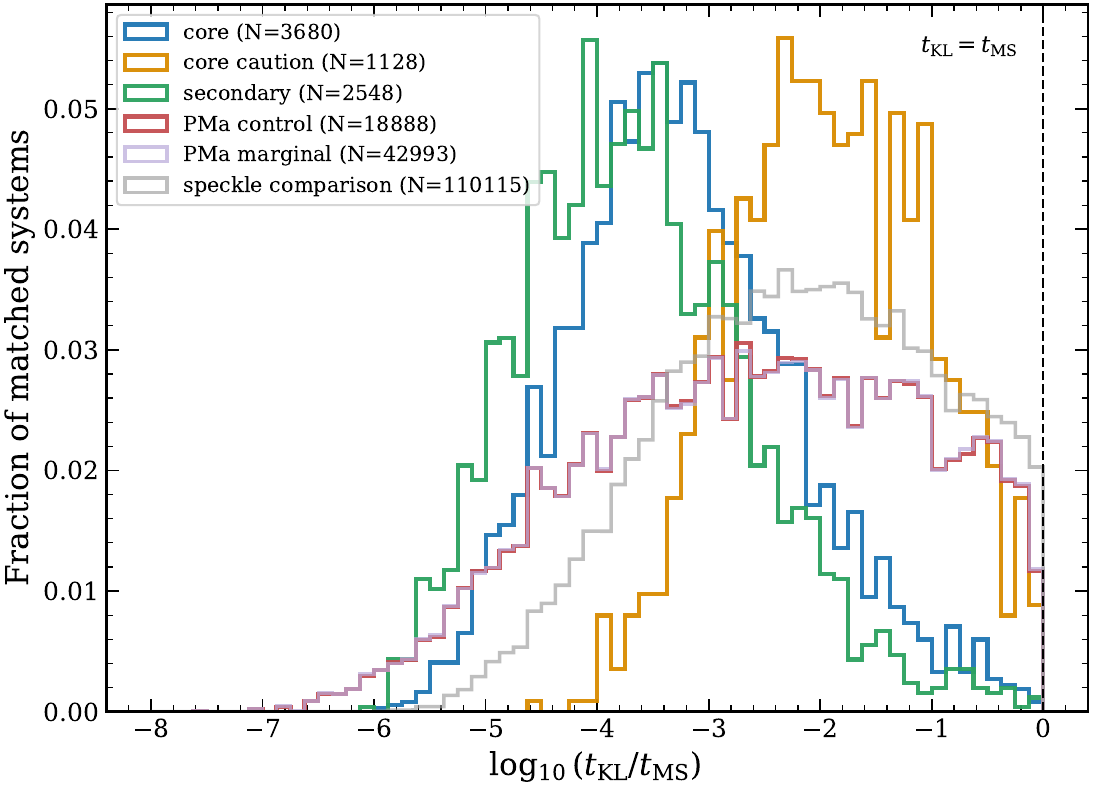}
    \caption{
    Distribution of normalized KL timescales for the matched
    synthetic configurations, separated by observational class. The vertical
    dashed line marks $t_{\rm KL}=t_{\rm MS}$. Systems to the left of this line
    can undergo secular evolution within the main-sequence lifetime of the Be
    primary. The histograms are normalized within each class to compare their
    relative distributions rather than their absolute sample sizes. The class labels correspond to the constraint groups described in the text.
    }
    \label{fig:tkl_by_class}
\end{figure}

Figure~\ref{fig:tkl_by_class} shows the distribution of the normalized
KL timescale for the matched configurations, separated by
observational class. The robust core sample corresponds to systems with the
most restrictive and reliable combinations of constraints, such as
period+speckle cases. The core-caution sample contains systems with useful but
less secure astrometric information. The secondary sample represents additional
matched configurations that satisfy the dynamical filters but are less
constraining observationally. The PMa-control and PMa-marginal samples trace
systems selected primarily through PMa information, while
the speckle-comparison sample traces the broader population of
speckle-detectable synthetic tertiaries. The vertical dashed line marks
$t_{\rm KL}=t_{\rm MS}$; configurations to the left of this line have secular
timescales shorter than the adopted main-sequence lifetime scale. A complementary view of the dependence of the secular timescale on the
synthetic hierarchy ratio and outer eccentricity is provided in
Appendix~\ref{sec:ap-hierarchy-timescale}.

The different classes occupy overlapping but not identical ranges of
$t_{\rm KL}/t_{\rm MS}$. The most constrained systems tend to select compact
inner architectures and therefore shorter secular timescales, whereas the
comparison samples span a broader range. This behaviour is expected because the
period, PMa, and speckle constraints restrict different combinations of
$a_{\rm in}$, $a_{\rm out}$, and projected separation. The important result is
that the accepted configurations generally satisfy the basic dynamical
requirement for secular evolution, defined in
Eq.~(\ref{eq:tkl_condition}).

\subsection{Outer separation and secular efficiency}
\label{subsec:sep_tkl}

\begin{figure}[!t]
    \centering
    \includegraphics[width=\columnwidth]{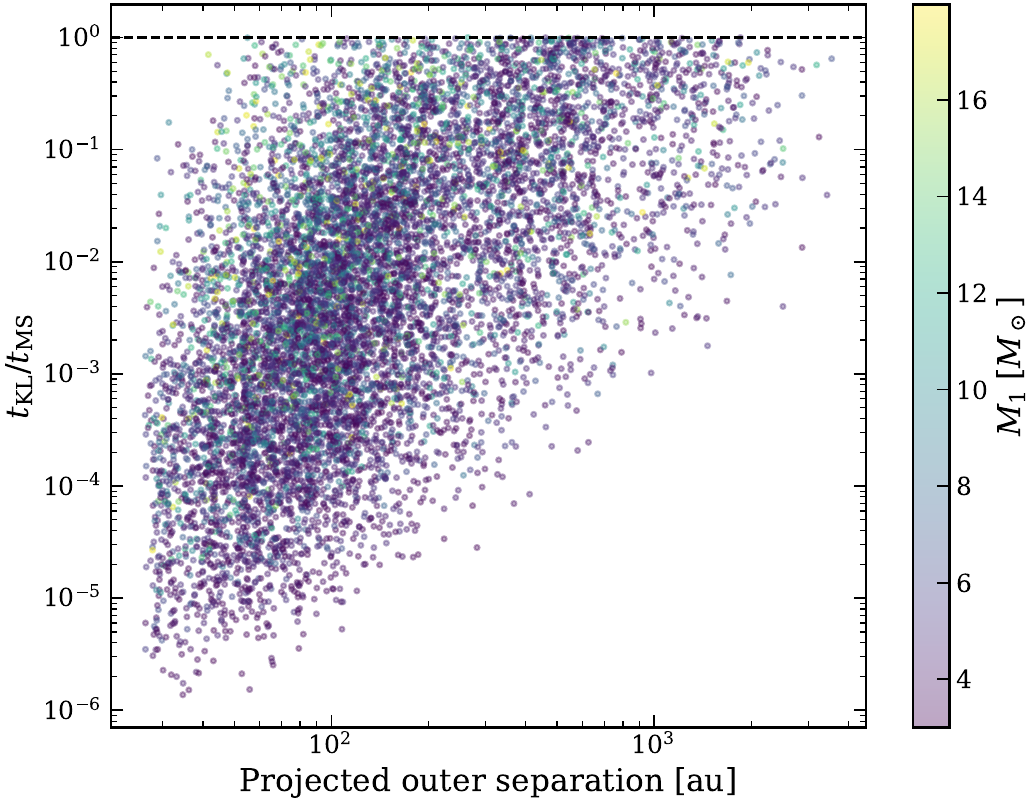}
    \caption{
    Projected physical separation of the synthetic outer companion as a
    function of normalized KL timescale. The colour scale indicates
    the primary mass. More compact projected separations generally correspond
    to shorter secular timescales, whereas wider companions tend to approach or
    exceed $t_{\rm KL}/t_{\rm MS}=1$. This figure illustrates the dynamical
    relevance of the speckle-resolved separation range.
    }
    \label{fig:sep_tkl}
\end{figure}

The connection between outer separation and secular efficiency is shown in
Fig.~\ref{fig:sep_tkl}. The projected physical separation of the outer
companion is plotted against $t_{\rm KL}/t_{\rm MS}$ for the accepted synthetic
systems. More compact projected separations generally correspond to shorter
KL timescales, while wider companions tend to approach or exceed the
$t_{\rm KL}=t_{\rm MS}$ limit. This trend illustrates why subarcsecond
companions detected by speckle interferometry can be dynamically relevant:
although they are too wide to explain short-period spectroscopic variability
directly, they may occupy the range of separations required to perturb an
unresolved inner binary.

\subsection{Consistency with the observed speckle separations}
\label{subsec:rho_consistency}

\begin{figure}[!t]
    \centering
    \includegraphics[width=\columnwidth]{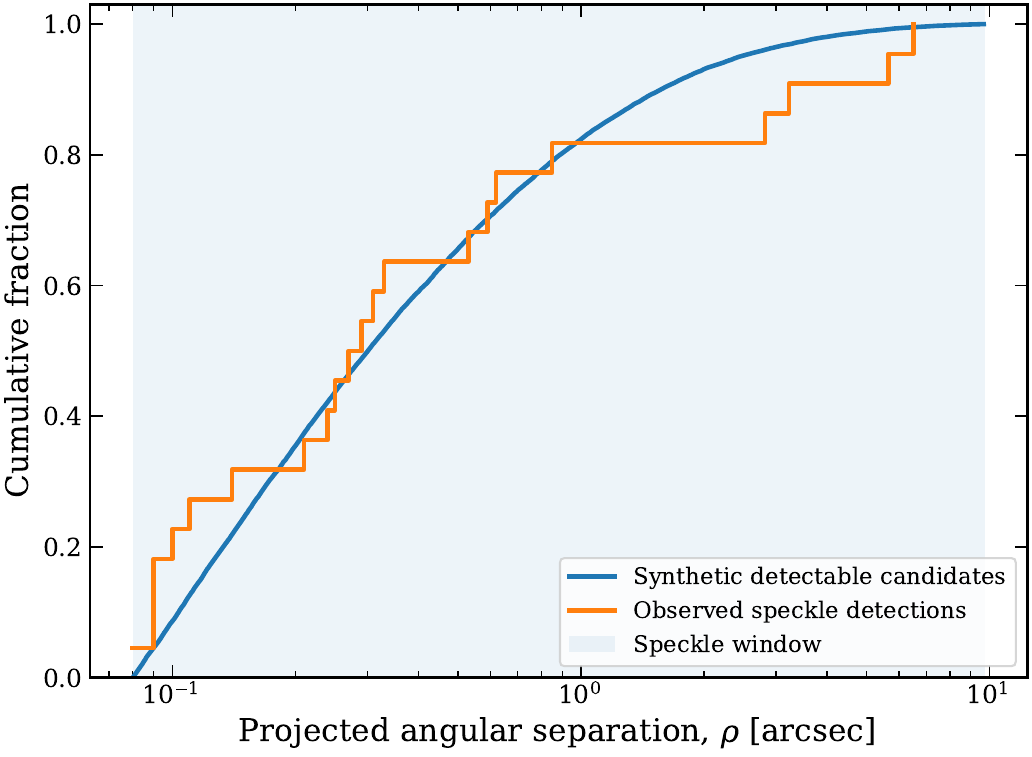}
    \caption{
    Cumulative distribution of projected angular separations for the synthetic
    detectable candidates compared with the observed speckle detections. The
    shaded region marks the adopted speckle-detection window. The comparison
    shows that the observed companions occupy the angular-separation range in
    which the synthetic secularly relevant companions are detectable. 
    The observed distribution contains individual speckle separations instead of unique targets; in the adopted inventory, 22 speckle separations correspond
to 20 unique systems, with HD~161004 and HD~76838 each contributing two resolved
pairs.
    }
    \label{fig:rho_cdf}
\end{figure}

We next compared the projected angular separations of the synthetic systems with
the observed speckle detections. Figure~\ref{fig:rho_cdf} compares the
cumulative distribution of the synthetic detectable candidates with the
observed speckle separations. The observed companions lie within the same
subarcsecond-to-few-arcsecond range occupied by the detectable synthetic
systems, supporting the interpretation that the observed companions can sample
the outer-orbit regime relevant for secular perturbations.

\begin{figure}[!t]
    \centering
    \includegraphics[width=\columnwidth]{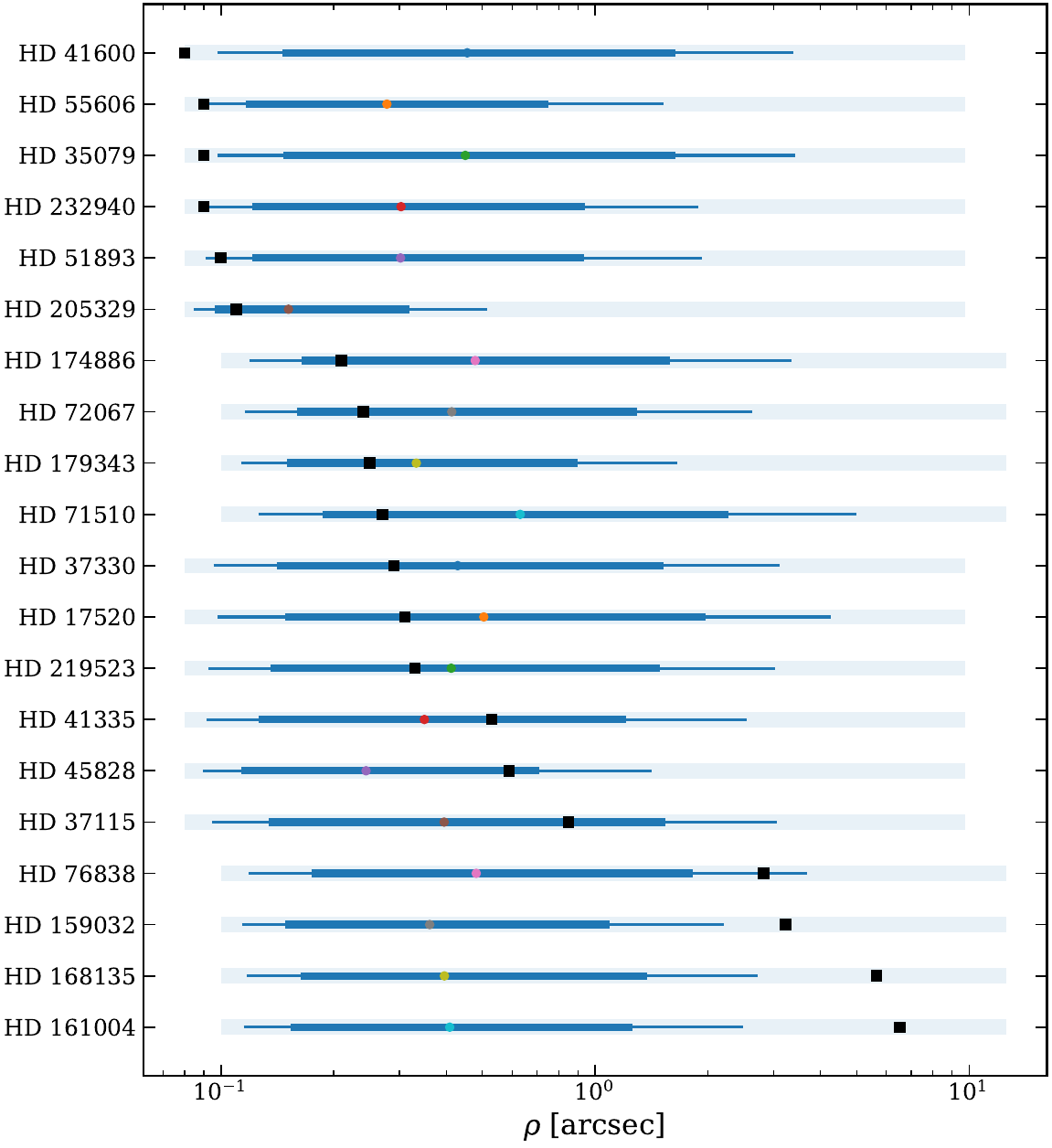}
    \caption{
    Comparison of observed speckle separations and projected separations
    of the accepted synthetic systems. Black squares mark the observed
    $\rho_{\rm obs}$ values, coloured points the model medians, thick bars
the 16--84 percentile ranges, and thin bars the 5--95 percentile
    ranges. The faint band indicates the adopted speckle-detection window. The
    full target-by-target distributions are shown in
    Appendix~\ref{sec:ap-rho-distributions}.
    }
    \label{fig:rho_summary}
\end{figure}

A target-by-target summary is shown in Fig.~\ref{fig:rho_summary}. For each
object, the accepted model sample was constructed by applying the dynamical
filters and the object-specific observational constraints. Therefore, this
figure is a consistency check rather than an unconstrained prediction of the
separation distribution. The full projected-separation distributions for each target are shown in Appendix~\ref{sec:ap-rho-distributions}.

\subsection{Target-by-target constraint matching}
\label{subsec:target_matching}

\begin{figure}[!t]
    \centering
    \includegraphics[width=\columnwidth]{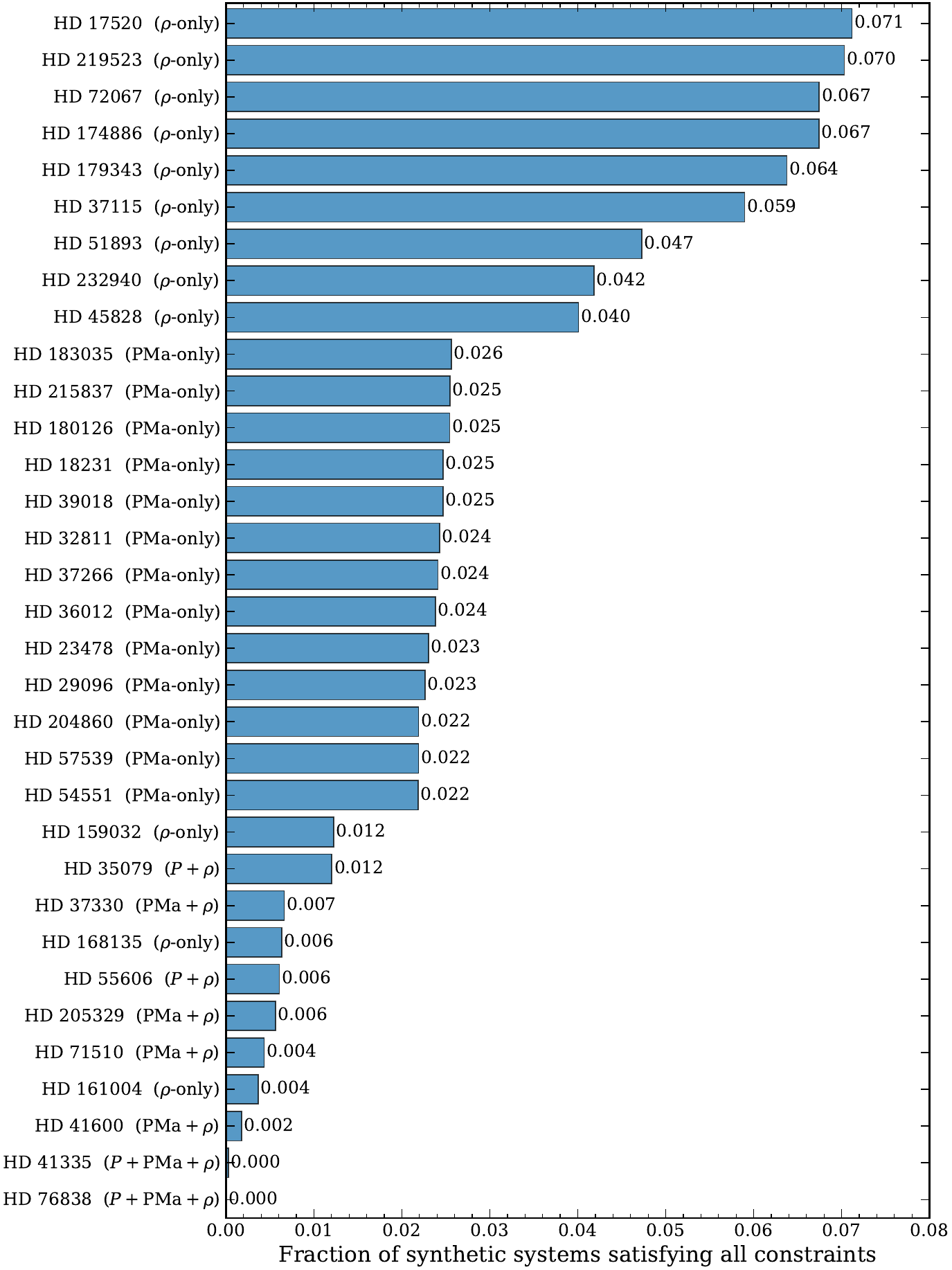}
\caption{
Fraction of synthetic systems satisfying the adopted
observational and dynamical constraints in the global
distance-aware experiment. The labels identify the
constraint class assigned to each target. PMa-only and
$\rho$-$\mathrm{only}$ systems provide comparison samples,
whereas combined constraints produce more selective
matches. These fractions describe the global experiment
and are distinct from the target-specific mass-aware
results in Table~\ref{tab:massaware_refined_results}.
}
    \label{fig:match_fraction}
\end{figure}

The fraction of synthetic systems satisfying all constraints for each target is
shown in Fig.~\ref{fig:match_fraction}. The comparison samples with only one
strong constraint, such as $\rho$-$\mathrm{only}$ or PMa-only systems, generally have
larger matched fractions. This is expected because fewer independent conditions
must be satisfied. In contrast, systems with combined constraints are more
restrictive. The smallest matched fractions occur when the model is required to
satisfy inner-period, PMa, and speckle constraints simultaneously.

The refined mass-aware results for the best-constrained
systems are summarized in Table~3.
In the refined simulations, HD~35079 and HD~55606
yield matched fractions of $8.79\times10^{-3}$ and
$4.08\times10^{-3}$, respectively, under their
combined period and speckle constraints. For
HD~205329, the PMa+speckle scenario yields
$f_{\rm match}=1.83\times10^{-3}$, with its
interpretation remaining dependent on the approximate
PMa orbital scale. HD~41335 admits separate
$P+\rho$ and PMa+$\rho$ solutions, with matched
fractions of $9.30\times10^{-4}$ and
$2.24\times10^{-3}$, respectively, but no accepted
configurations when all three constraints are imposed
simultaneously.

The accepted configurations have median
$t_{\rm KL}/t_{\rm MS}<1$ and satisfy the adopted
interaction proxy. Their median inner-periastron
distances are comparable to or smaller than the
combined stellar radii, supporting the dynamical
plausibility of secularly driven close-binary
interaction under the adopted assumptions.

\begin{table*}[ht!]
\caption{Refined dynamical results for the key target-specific scenarios.}
\label{tab:massaware_refined_results}
\centering
\begin{tabular}{lcccccc}
\hline\hline
Target & Constraints & $f_{\rm match}$ & $t_{\rm KL}/t_{\rm MS}$ &
$a_{\rm in}$ & $\rho_{\rm model}$ &
$r_{\rm peri,in}/(R_1+R_2)$ \\
       &             &                 &                         &
(au)   & (arcsec)    &                         \\
\hline
HD~35079 & $P+\rho$ &
0.008787 &
$0.0091^{+0.0457}_{-0.0073}$ &
$0.168^{+0.0256}_{-0.0211}$ &
$0.0853^{+0.0196}_{-0.0159}$ &
$0.615^{+1.34}_{-0.553}$ \\

HD~55606 & $P+\rho$ &
0.004083 &
$0.0285^{+0.103}_{-0.0227}$ &
$0.750^{+0.113}_{-0.093}$ &
$0.0860^{+0.0194}_{-0.0160}$ &
$0.751^{+1.39}_{-0.664}$ \\

HD~205329 & PMa+$\rho$ &
0.001827 &
$0.380^{+0.416}_{-0.261}$ &
$2.32^{+0.46}_{-0.50}$ &
$0.0961^{+0.0273}_{-0.0139}$ &
$0.689^{+1.34}_{-0.617}$ \\

HD~41335 & $P+\rho$ &
0.000930 &
$0.508^{+0.319}_{-0.305}$ &
$0.897^{+0.164}_{-0.130}$ &
$0.451^{+0.118}_{-0.061}$ &
$0.659^{+1.42}_{-0.553}$ \\

HD~41335 & PMa+$\rho$ &
0.002237 &
$0.332^{+0.436}_{-0.250}$ &
$2.59^{+0.55}_{-0.53}$ &
$0.464^{+0.120}_{-0.069}$ &
$0.719^{+1.24}_{-0.652}$ \\

HD~41335 & $P+\mathrm{PMa}+\rho$ &
$0.000000$ &
-- &
-- &
-- &
--\\
\hline
\end{tabular}
\tablefoot{
The table gives the results of the refined mass-aware Monte Carlo runs.
The matched fraction ($f_{\rm match}$) is the fraction of generated systems
satisfying the dynamical filters and the target-specific observational
constraints. Values with uncertainties are medians with 16--84 percentile
ranges. 
For HD~41335, the simultaneous
$P+\mathrm{PMa}+\rho$ constraint yields
$f_{\rm match}=0$ (no accepted configurations)
at the fiducial tolerance $f=0.3$.
}
\end{table*}

These results define a hierarchy of dynamical plausibility among the refined
systems. The matched fraction should be interpreted together with the
observational quality of the constraints: larger values indicate combinations
of observables that are easier to satisfy within the adopted synthetic
population, whereas smaller values correspond to more selective or more
tensioned cases. In this sense, HD~35079 and HD~55606 remain the strongest
systems not because they have the largest matched fractions, but because their
independently measured inner periods and speckle separations are reproduced
simultaneously by stable, secularly active hierarchical triples. HD~205329 is a
dynamically plausible but astrometrically cautionary case, while HD~41335 is a
diagnostic system in which the period and PMa constraints likely trace
different dynamical scales.

\subsection{Inner-orbit constraints}
\label{subsec:inner_constraints}

\begin{figure}[!t]
    \centering
    \includegraphics[width=\columnwidth]{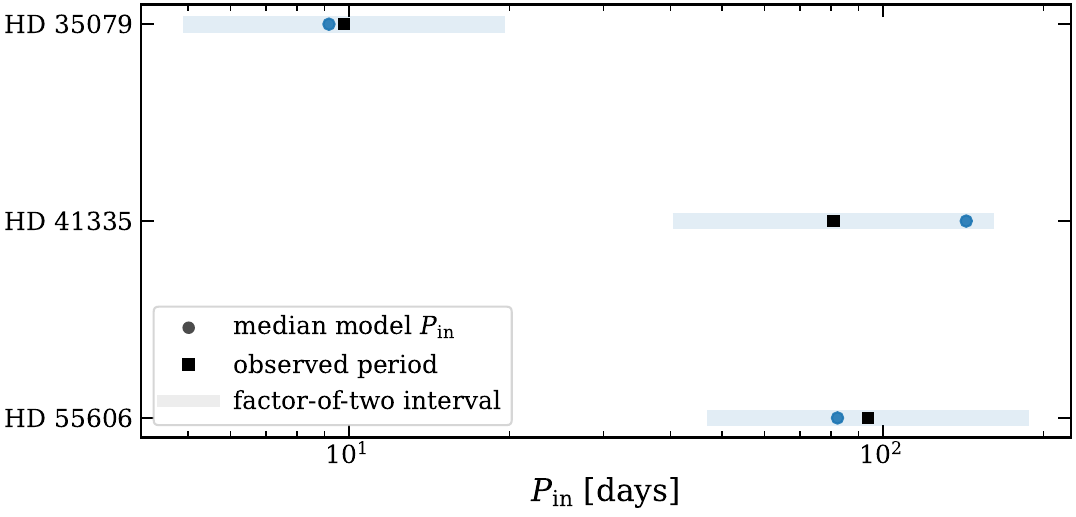}
    \caption{
    Comparison of observed and synthetic inner periods for systems with
    available period constraints. The shaded region indicates a
    factor-of-two interval around the observed value. These systems provide the
    strongest constraints on the inner binary scale.
    }
    \label{fig:period_summary}
\end{figure}

For systems with reported or inferred inner periods, the accepted synthetic
models must reproduce the observed period within the adopted tolerance. Figure~\ref{fig:period_summary} compares the observed and modelled inner
periods for these systems. The period constraint is particularly important
because it fixes the inner semi-major axis through Kepler's law and therefore
affects both the KL timescale and the periastron scale reached after secular
eccentricity excitation.

\begin{figure}[!t]
    \centering
    \includegraphics[width=\columnwidth]{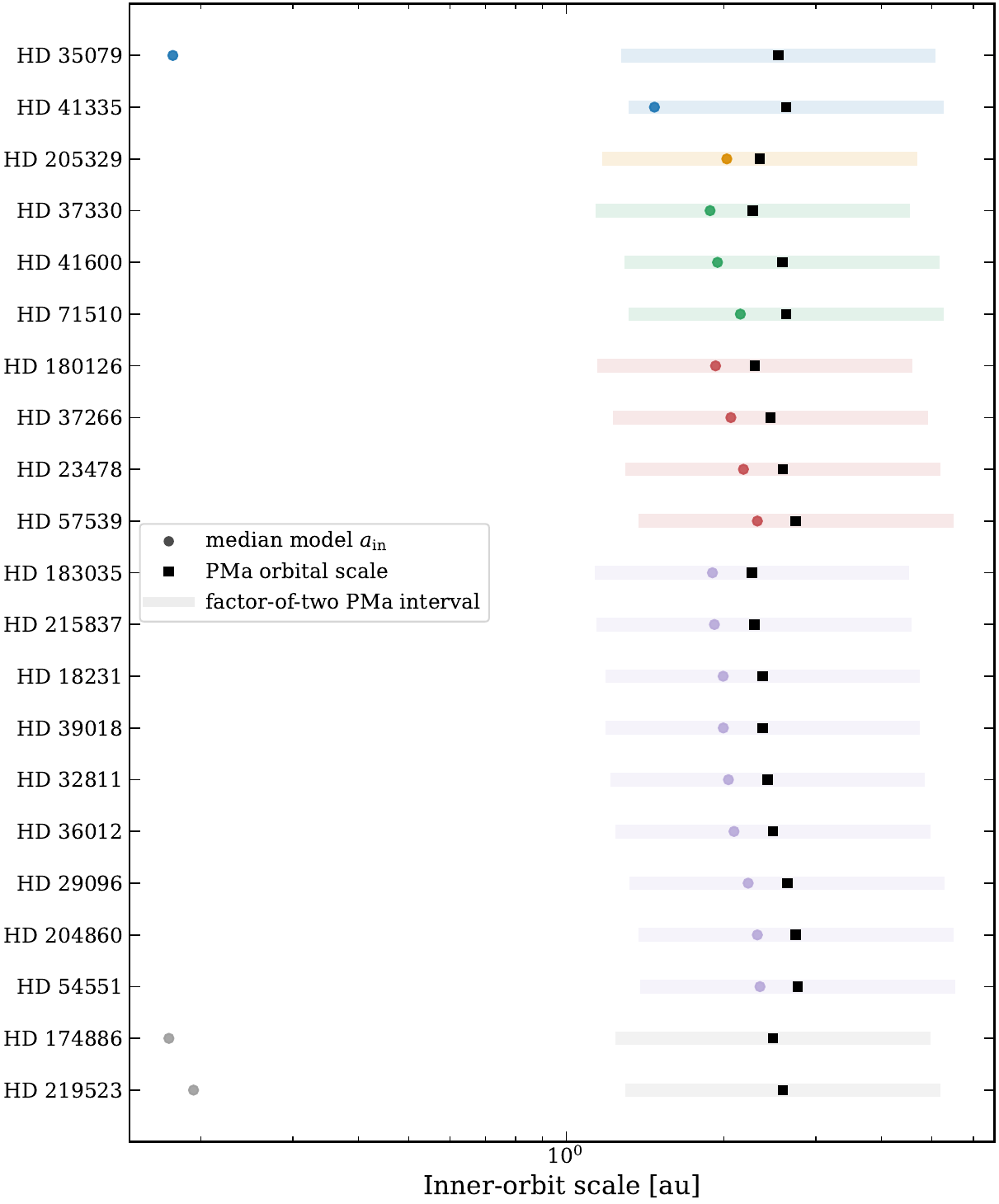}
    
    \caption{
Inner-orbit scale of the accepted synthetic systems
compared with the PMa-inferred orbital scale.
Coloured circles indicate the median model inner
semi-major axes, with colours identifying the
observational classes used in the synthetic experiment.
Black squares indicate the PMa-inferred orbital scales,
while the faint horizontal bars represent the adopted
matching intervals (see Sect.~\ref{subsec:target_classes}). 
}
    \label{fig:inner_scale}
\end{figure}

For systems with PMa information, the PMa-inferred orbital scale provides an
independent constraint on unresolved multiplicity. Figure~\ref{fig:inner_scale}
compares the PMa scale with the accepted synthetic inner semi-major axis. The
PMa constraint is useful but should be interpreted with caution because it is
not a full orbital solution and can be affected by projection, luminosity
ratios, astrometric systematics, or higher-order multiplicity.

\section{Discussion}
\label{sec:discussion}

The results above show that several Be-star multiple systems can be embedded
within dynamically stable hierarchical architectures capable of secular
evolution on timescales shorter than the main-sequence lifetime of the Be
primary. The synthetic sampling of mutual
inclinations identifies the KL-active architectures that are simultaneously
compatible with the observed separations, periods, and/or PMa-inferred scales.
These configurations therefore define the dynamically plausible region for
triple-assisted interaction. 
This does not imply that KL cycles directly produced the
present-day Be disks. Rather, it establishes a more limited but physically
important result: the observed architectures can satisfy the dynamical
conditions required for a tertiary companion to drive an unresolved inner
binary towards the interaction regime.

\subsection{Dynamical plausibility of triple-assisted interaction}
\label{subsec:discussion_dynamical_plausibility}

The key requirement for the proposed channel is that the outer companion should
not be merely a wide visual neighbour. It must occupy a range of separations
where secular perturbations can act within the lifetime of the Be primary. The
matched configurations satisfy this condition in two ways. First, the
KL timescale is generally shorter than $t_{\rm MS}$, implying that
secular evolution is dynamically available. Second, the secularly excited inner
periastron can approach the interaction proxy defined in
Eq.~(\ref{eq:interaction_proxy}) for the best-constrained systems.

\begin{figure}[!t]
    \centering
    \includegraphics[width=\columnwidth]{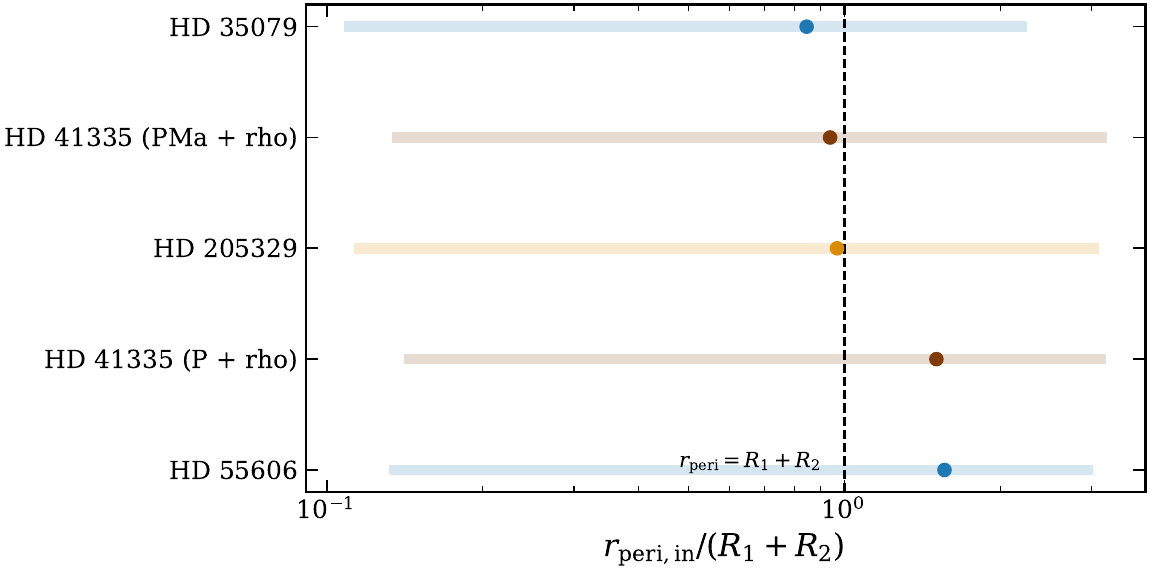}
    \caption{
    Inner periastron distance normalized by the sum of the stellar radii for
    the refined dynamically matched scenarios. The vertical dashed line marks
    $r_{\rm peri,in}=R_1+R_2$. Points and horizontal intervals show the median
    and percentile ranges of the accepted synthetic models. Values near unity
    indicate systems in which KL eccentricity excitation can bring the inner
    binary to the adopted interaction scale.
    }
    \label{fig:inner_periastron}
\end{figure}

This point is summarized in Fig.~\ref{fig:inner_periastron}, which shows the
inner periastron distance normalized by the combined stellar radii. Systems
with values near unity are those in which KL eccentricity excitation can bring
the inner binary close to the stellar-interaction scale. These configurations
are therefore the most relevant for triple-assisted interaction, even though
the present calculation does not follow the subsequent binary evolution.

The systems with period-plus-speckle constraints provide the cleanest test of
this scenario because the inner and outer orbital scales are constrained
independently. In this sense, HD~35079 and HD~55606 are the most robust
dynamical cases in the present sample. Their observed inner periods and
speckle separations can be simultaneously reproduced by stable, secularly
active triples. HD~205329 is also dynamically plausible, but its interpretation
depends on the PMa constraint and is therefore more sensitive to astrometric
systematics and unresolved higher-order structure.

The PMa-plus-speckle systems HD~41335, HD~37330, HD~41600, HD~205329, and
HD~71510 provide a complementary set of observational candidates because they
combine a resolved visual companion with astrometric evidence of unresolved
multiplicity. Their interpretation is more cautionary because the PMa scale is
an approximate astrometric constraint. These systems are therefore natural targets
for follow-up astrometry, spectroscopy, optical interferometry, and
spectro-interferometry. In particular, they may represent strong observational
hierarchical-triple candidates even when they do not emerge as the strongest
KL-active matches under the specific inclination and interaction criteria
adopted in this work.

In this interpretation, the speckle-resolved companion is not the direct mass
donor. Its role is dynamical: it can act as the tertiary perturber that drives
the unresolved inner binary towards high eccentricity and small periastron
distances \citep{2016ComAC...3....6T}. The subsequent Be-star spin-up would then occur through interaction
inside the unresolved inner pair, either by mass transfer onto the
Be component or through a merger-like event. Thus, the triple-assisted channel
connects the observed wide companion to Be-star formation indirectly, through
its secular effect on the inner orbit.

\subsection{The special case of HD~41335}
\label{subsec:discussion_hd41335}

HD~41335 (V696~Mon) is a particularly interesting case.
The complex photometric and spectroscopic behaviour
discussed by \citet{2026A&A...712A.226B}, including
the difficulties in interpreting its orbital variability,
makes this system especially relevant for investigating
the possible role of tertiary-driven secular evolution.
Our simulations show that the period-plus-speckle and
PMa-plus-speckle constraints independently admit stable,
KL-active hierarchical configurations.
Moreover, relaxing the common matching tolerance from
$f=0.3$ to $f=0.5$ reveals a small population of
configurations ($f_{\rm match}\sim10^{-4}$) satisfying
all three constraints, with inner and outer orbital scales
comparable to those inferred observationally.
These results highlight HD~41335 as a promising target
for investigating a potentially more complex hierarchical
architecture, in which the spectroscopic, astrometric,
and speckle signatures provide complementary information
on the system's dynamical evolution.

\begin{figure}[!t]
    \centering
    \includegraphics[width=\columnwidth]{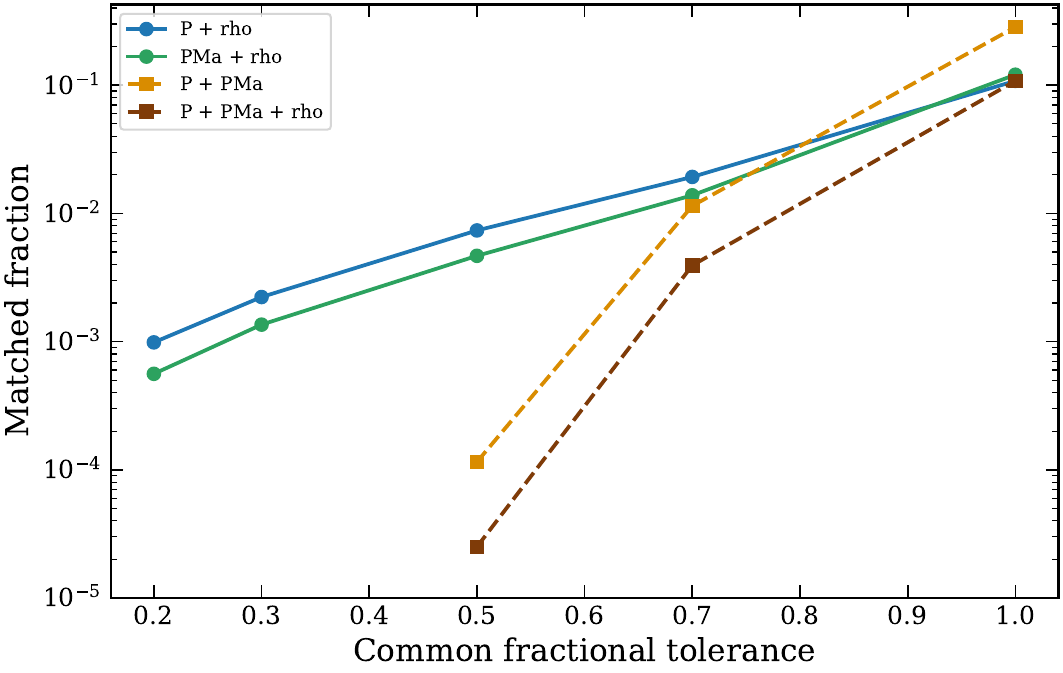}
    \caption{
    Sensitivity of the matched fraction for HD~41335 to the adopted common
fractional tolerance. Each curve corresponds to a different combination of
constraints: inner period ($P$), PMa-inferred orbital scale, and speckle
separation ($\rho$). The combinations $P+\rho$ and PMa+$\rho$ remain viable
over a broad range of tolerances, whereas combinations involving both $P$ and
PMa are strongly suppressed. This indicates that
    the observed period and PMa scale are unlikely to describe the same simple
    inner orbit in the adopted model.
    }
    \label{fig:hd41335_sensitivity}
\end{figure}

The sensitivity analysis in Fig.~\ref{fig:hd41335_sensitivity}
provides further insight into this result by showing how
the different observational constraints interact as the
common matching tolerance is varied.
The $P+\rho$ and PMa+$\rho$ combinations remain viable
over a broad range of tolerances, whereas their simultaneous
application is considerably more selective.
The emergence of configurations satisfying all three
constraints at larger tolerances suggests that the observed
period, PMa signal, and speckle companion probe different
aspects of a complex hierarchical architecture.

\subsection{Angular-momentum plausibility}
\label{subsec:discussion_spinup}

The simulations show that the observed architectures can reach the interaction
regime, but they do not compute the mass transferred, the angular momentum
retained by the Be star, or the subsequent spin evolution. Addressing this
stage requires binary-evolution calculations including tides, mass transfer,
angular-momentum loss, stellar structural evolution, and possible merger
physics, which are beyond the scope of this analysis.

As a complementary consistency check, we estimated
the angular momentum required to spin up the Be star,
following the classical argument of
\citet{1981A&A...102...17P} and binary-interaction
models of rapidly rotating massive stars
\citep{2013ApJ...764..166D}.
This angular-momentum estimate is intentionally
decoupled from the dynamical Monte Carlo filtering.
The secular calculations determine whether the observed
architectures can bring the unresolved inner binary into an interaction
regime. The spin-up estimate then asks a separate question: if such an
interaction supplies material and angular momentum to the Be
component, is the required accreted mass fraction prohibitively large? Thus, $\Delta M_{\rm req}$ (see Eq.~\ref{eq:spinup_budget}) is evaluated after the dynamical filtering as a
post-processing consistency check that connects the accepted architectures with
the angular-momentum scale required for Be-star spin-up.

This estimate provides a useful order-of-magnitude benchmark because it
quantifies, in the context of the accepted dynamical scenarios, the classical
expectation that material added with high specific angular momentum can
produce substantial spin-up after only modest fractional mass growth. It
therefore provides a needed reference scale for evaluating whether the
interaction regime identified by the dynamical simulations could plausibly
supply the angular-momentum scale needed for the Be phenomenon.

If the Be component accretes material from the
unresolved inner companion, or if the inner interaction proceeds through a
merger-like event, the delivered angular momentum can spin up the Be star. We
parameterize the specific angular momentum of the material added to the Be
primary as $j_{\rm acc}$ and define its efficiency relative to the Keplerian
value at the stellar surface as
\begin{equation}
    \eta_j=\frac{j_{\rm acc}}{j_{\rm Kep}}.
    \label{eq:eta_j}
\end{equation}

We parameterize the specific angular momentum of the material added to the Be primary 
as $j_{\rm acc}$ whereas $j_{\rm Kep}=(GMR)^{1/2}$ is the Keplerian specific angular momentum at the 
equator of a star of mass $M$ and radius $R$. With this, we define the specific-angular-momentum 
delivery efficiency, $\eta_j$, normalized to the Keplerian value at the stellar surface through Eq.~\ref{eq:eta_j}.
This normalization is
appropriate for disk-like or equatorially delivered material, as expected in
efficient spin-up scenarios. Less equatorial or less efficiently retained
angular-momentum delivery is represented phenomenologically by smaller values
of $\eta_j$. We also define the normalized
stellar rotation rate as
\begin{equation}
    \omega=\frac{\Omega}{\Omega_{\rm crit}},
    \label{eq:omega_norm}
\end{equation}
where $\Omega$ is the stellar angular velocity and $\Omega_{\rm crit}$ is the
critical angular velocity. Thus, $\omega_0=\Omega_0/\Omega_{\rm crit}$ and
$\omega_f=\Omega_f/\Omega_{\rm crit}$ denote the initial and final normalized
rotation rates, respectively. With these definitions, the fractional mass
required to spin a star from $\omega_0$ to $\omega_f$ can be written, to order
of magnitude, as
\begin{equation}
    \frac{\Delta M_{\rm req}}{M}
    \simeq
    \frac{k^2}{\eta_j}
    \left(\omega_f-\omega_0\right),
    \label{eq:spinup_budget}
\end{equation}
where $k^2$ is the dimensionless squared radius of gyration, defined through
the stellar moment of inertia $I=k^2MR^2$. In this expression,
$\Delta M_{\rm req}$ is the amount of mass that must be added to the Be component
to provide the angular momentum needed to change the stellar rotation from
$\omega_0$ to $\omega_f$. Equation~(\ref{eq:spinup_budget}) is used as a perturbative
angular-momentum estimate, appropriate for evaluating the spin-up scale
associated with modest fractional mass growth. In this approximation, the
stellar mass, radius, moment-of-inertia coefficient $k^2$, and critical angular
velocity are treated as representative values during the angular-momentum
delivery.

The parameters $k^2$, $\omega_0$, $\omega_f$, and $\eta_j$ are not measured
for the individual systems. They are sampled over representative ranges in
order to estimate the angular-momentum scale associated with Be-star spin-up.
The resulting $\Delta M_{\rm req}$ values should therefore be read as
order-of-magnitude spin-up budgets for the Be component, rather
than as reconstructed mass-transfer histories.

Nevertheless, once the inner binary is dynamically driven towards interaction,
it is useful to ask whether the angular-momentum budget required for rapid
rotation is prohibitive. We therefore used the order-of-magnitude relation
defined in Eq.~(\ref{eq:spinup_budget}) as a global plausibility estimate.
This expression follows the classical spin-up argument that accretion with high
specific angular momentum can drive a star towards critical rotation after
accreting only a modest fraction of its mass \citep{1981A&A...102...17P}. It is
also consistent with binary-population models in which mass transfer and
mergers produce a substantial population of rapidly rotating massive stars
\citep{2013ApJ...764..166D}.

\begin{figure*}[!t]
    \centering
    \includegraphics[width=\textwidth]{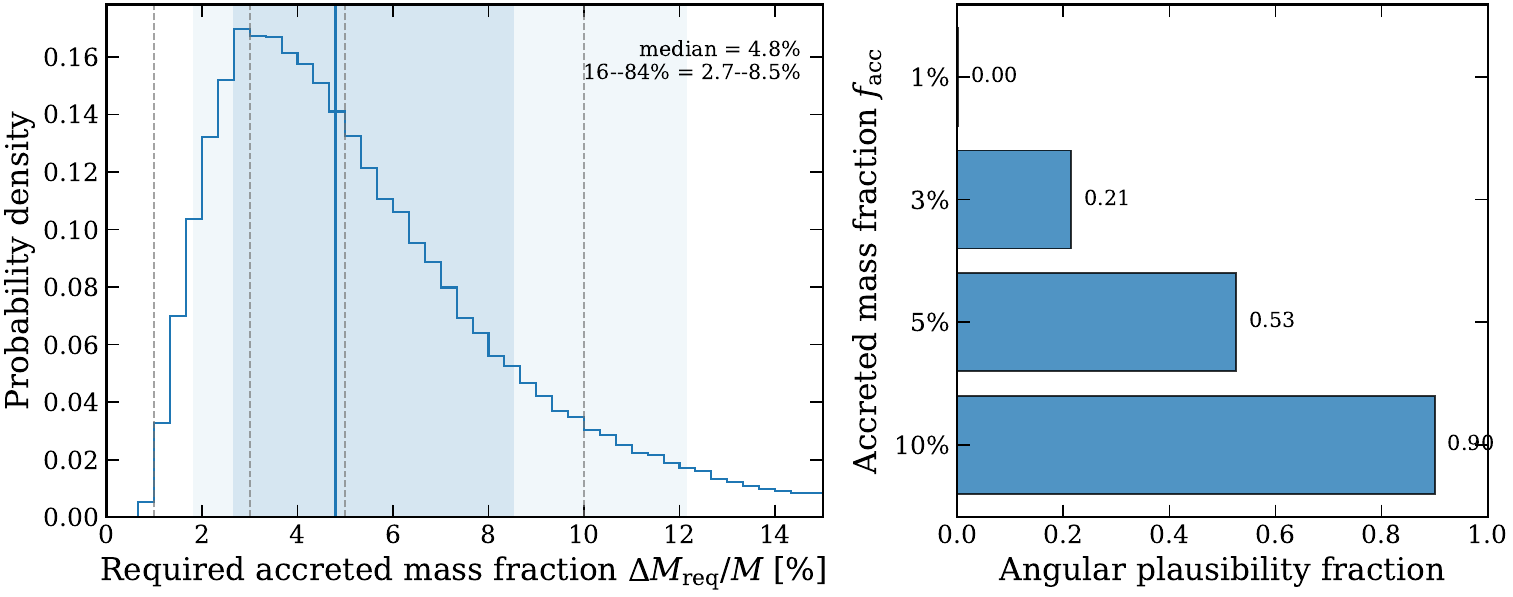}
    \caption{
    Global angular-momentum plausibility budget for spinning up a Be star.
    Left: Probability density of the required accreted mass fraction,
    $\Delta M_{\rm req}/M$, computed from Eq.~(\ref{eq:spinup_budget}) for
    $k^2=0.03$--0.10, $\omega_0=0.2$--0.7,
    $\omega_f=0.9$--1.0, and $\eta_j=0.3$--1.0. The vertical solid line marks
    the median, and the shaded bands indicate representative percentile ranges.
    Right: Fraction of realizations for which a given available accreted mass
    fraction, $f_{\rm acc}$, is sufficient to reach near-critical rotation. 
    }
    \label{fig:spinup_budget}
\end{figure*}

Figure~\ref{fig:spinup_budget} summarizes the global angular-momentum
plausibility estimate. The left panel shows the distribution of the required effective mass fraction, $\Delta M_{\rm req}/M$, where $M$ is the mass of the present-day Be component
that is being spun up, calculated by sampling the
allowed ranges of $k^2$, $\omega_0$, $\omega_f$, and $\eta_j$ in
Eq.~(\ref{eq:spinup_budget}). The median value,
$\Delta M_{\rm req}/M\simeq 4.8\%$, is a key result of this calculation:
under the adopted angular-momentum assumptions, spinning a Be star from a
moderate initial rotation rate to near-critical rotation does not require
accretion of a large fraction of the stellar mass. The 16--84 percentile range,
approximately $2.7$--$8.5\%$, shows that this conclusion remains valid across
the explored range of stellar structure and angular-momentum delivery
parameters. It should therefore
be interpreted as the effective amount of high-angular-momentum material, or
equivalent merger-delivered angular momentum, required to spin up the observed
Be star. For a low-mass donor, the same absolute requirement could correspond
to a substantial fraction of the donor mass, so the values in
Table~\ref{tab:individual_spinup_budget} should be read as angular-momentum
budget estimates rather than conservative mass-transfer predictions.

The right panel provides a complementary way to read the same result. It shows
the fraction of Monte Carlo realizations for which an assumed available
accreted mass fraction, $f_{\rm acc}$, is sufficient to satisfy the spin-up
requirement. If only $f_{\rm acc}=3\%$ of the primary mass is effectively
accreted with the adopted angular-momentum efficiencies, about 21 per cent of
the realizations reach the required budget. For $f_{\rm acc}=5\%$, this
fraction rises to approximately 53 per cent, and for $f_{\rm acc}=10\%$ it
reaches about 90 per cent. Thus, the figure shows that once a secularly driven
interaction is dynamically allowed, the angular-momentum budget required for
near-critical rotation is achievable for physically modest accreted mass
fractions.

The simulations allow this fractional requirement to be translated
into an illustrative absolute accreted mass for the key systems with reported masses, listed in Table~\ref{tab:individual_spinup_budget}. For the two robust
period-plus-speckle cases, the median angular-momentum budgets correspond to
$\Delta M_{\rm req}\simeq0.24\,M_\odot$ for HD~35079 and
$\simeq0.30\,M_\odot$ for HD~55606, with 16--84 percentile ranges of
$0.13$--$0.42\,M_\odot$ and $0.16$--$0.53\,M_\odot$, respectively. For the
cautionary PMa-plus-speckle case HD~205329, the corresponding median value is
$\simeq0.53\,M_\odot$. For HD~41335, the two viable partial constraint
combinations yield larger absolute requirements, $\simeq0.74$--$0.79\,M_\odot$,
because of the higher adopted primary mass. No value is reported for the
combined $P+$PMa+$\rho$ scenario because no accepted configurations are found
under the adopted tolerances.

These values provide an illustrative, system-scaled angular-momentum budget.
They are obtained by multiplying the global fractional spin-up requirement by
the median mass of the Be component in each accepted sample,
thereby translating the global estimate into the physical mass scale of each
scenario. Under the assumptions adopted here, accretion or merger-like delivery
of a few per cent of that mass is sufficient in a large fraction of
realizations. The result therefore supports the plausibility of near-critical
rotation following secularly driven interaction, while leaving the detailed
spin evolution to future binary-evolution modelling.

A useful comparison case is $\nu$~Gem (HD~45542), a confirmed hierarchical
triple containing an outer Be star \citep{2021ApJ...916...24K}. Its nearly
coplanar architecture illustrates that multiplicity alone is not sufficient for
the triple-assisted interaction channel; the orbital configuration must also
allow efficient secular eccentricity excitation. This distinction is central
to the present analysis.

\begin{table}[ht!]
\caption{Illustrative individual angular-momentum budgets.}
\label{tab:individual_spinup_budget}
\centering
\begin{tabular}{lcc}
\hline\hline
Scenario & $M_1$ & $\Delta M_{\rm req}$ \\
         & ($M_\odot$) & ($M_\odot$) \\
\hline
HD~35079 &
$4.92$ &
$0.24^{+0.18}_{-0.11}$ \\

HD~55606 &
$6.20$ &
$0.30^{+0.23}_{-0.13}$ \\

HD~205329 &
$10.97$ &
$0.53^{+0.41}_{-0.24}$ \\

HD~41335 ($P+\rho$) &
$15.41$ &
$0.74^{+0.58}_{-0.33}$ \\

HD~41335 (PMa+$\rho$) &
$16.46$ &
$0.79^{+0.62}_{-0.35}$ \\
\hline
\end{tabular}
\tablefoot{
The absolute values are obtained by multiplying the global spin-up requirement,
$\Delta M_{\rm req}/M=4.8^{+3.7}_{-2.1}\%$, by the median mass of the
Be star in each accepted sample. They represent the effective
high-angular-momentum mass required to spin up the observed Be star, or the
equivalent angular-momentum delivery in a merger-like event, not the total mass
lost by the donor.
}
\end{table}

\subsection{Limitations of the model}
\label{subsec:discussion_limitations}

Several limitations should be kept in mind. Namely, the speckle separation is an
instantaneous projected separation, not the true outer semi-major axis. Although
the synthetic orbits are projected onto the sky before comparison with
$\rho_{\rm obs}$, the lack of full orbital coverage means that individual
systems remain degenerate in phase, inclination, eccentricity, and orientation.

The PMa constraint is not a full orbital solution. It provides evidence
of unresolved astrometric motion and an approximate orbital scale, but it can
be affected by photocentre motion, luminosity ratios, \textit{Gaia} astrometric quality,
and unresolved multiple components. For this reason, PMa-based systems are
interpreted as cautionary or diagnostic unless supported by independent
constraints.

The adopted main-sequence lifetime is based on a simple mass-lifetime scaling.
This approximation is sufficient for identifying clearly fast secular systems,
such as HD~35079 and HD~55606, whose accepted configurations have
$t_{\rm KL}/t_{\rm MS}\ll1$. The interpretation of more marginal cases,
especially HD~205329 and HD~41335, is more sensitive to the adopted lifetime
scale and should be revisited with detailed stellar-evolution tracks once more
precise stellar parameters are available.

The interaction proxy is intentionally simple. The condition
$r_{\rm peri,in}\leq f_{\rm int}(R_1+R_2)$ identifies configurations where KL
eccentricity excitation can bring the inner binary close to the stellar scale,
but it does not determine whether tides, mass transfer, common-envelope
evolution, or merger actually occurs. A full treatment would require coupling
secular dynamics to binary stellar evolution.

The present model is secular and therefore does not explicitly include
mean-motion resonances between the inner and outer orbits. This is appropriate
for the accepted configurations, which have large hierarchy ratios and are not
expected to be dominated by low-order orbital commensurabilities. Resonant
effects could become relevant for more compact or less hierarchical
architectures, but such systems are largely removed by the stability and
secular-regime filters adopted here.

The angular-momentum calculation is global rather than system-specific. The
parameters $k^2$, $\omega_0$, $\omega_f$, and $\eta_j$ are sampled from
representative ranges rather than measured for each target. The calculation therefore tests whether the spin-up budget is plausible in principle, not whether a particular
Be star was spun up by the modelled interaction.

The companion fractions discussed in this work quantify detections within the
speckle-accessible separation and contrast range. This makes them directly
comparable to other high-angular-resolution surveys over similar separation
ranges, such as the Very Large Telescope (VLT)/NACO study of \citet{2010MNRAS.405.2439O}, but
complementary to global OB-star multiplicity fractions, which combine
spectroscopic, interferometric, astrometric, and wide-field visual imaging techniques over a
much broader range of orbital scales \citep{2013ARA&A..51..269D,2017ApJS..230...15M}.
The relative numbers of binaries and higher-order multiples depend strongly on
primary mass and orbital scale; therefore, the systems considered here should
be interpreted as a dynamically selected subset rather than as a complete
multiplicity census.

Finally, the spin-up estimates presented in Sect.~\ref{subsec:discussion_spinup}
provide a global angular-momentum scale for the observed Be star. A
system-specific reconstruction would require binary-evolution calculations that
follow the transfer efficiency, angular-momentum retention, stellar-structure
response, and possible merger-like evolution.

\subsection{Future observational tests}
\label{subsec:discussion_future}

The most direct way to test the proposed scenario is to obtain additional
orbital information. Continued speckle monitoring can determine whether the
resolved companions share common proper motion and can eventually constrain
outer orbits. High-resolution spectroscopy can test for short-period inner
companions, line-profile variability, or signatures of stripped companions.
Improved astrometry from future \textit{Gaia} releases will also refine PMa
measurements and help distinguish photocentre effects from true orbital motion. 
Optical interferometry and spectro-interferometry can
provide complementary constraints on the geometry
of rapidly rotating stars, including their inclination
and projected rotation-axis orientation
\citep{2018MNRAS.480.1263H}. Combined with orbital astrometry
and spectroscopy, such measurements may help constrain
the three-dimensional architecture of candidate
hierarchical systems.

If the secular channel is currently active, it could leave observable
signatures through long-term changes in spectroscopic variability,
radial velocities, emission-line profiles, or disk-feeding episodes.
A past interaction could also leave residual circumstellar or
circumbinary material, particularly following non-conservative
mass transfer or a merger-like event. In such cases, only part of
the mass lost by the donor needs to be retained by the Be component
to provide the required angular momentum, while the remaining
material could be dispersed or temporarily stored in extended
structures \citep{2026MNRAS.552g1823M} influenced by the binary and tertiary orbits.
Although predicting the morphology and lifetime of this material
is beyond the scope of the present work, the dynamically plausible
systems identified here provide promising targets for searches
for extended emission, shell-like structures, infrared excess,
and long-term disk variability. Investigating these signatures will require repeated
speckle astrometry to trace outer-orbit motion,
time-domain spectroscopy to characterise inner
companions and line-profile variability, future \textit{Gaia}
astrometry, and optical interferometry or
spectro-interferometry to constrain orbital orientations
and, where feasible, mutual inclinations.

The highest-priority targets are those that combine a resolved speckle
companion with an independently constrained inner period or PMa signal. In this
sample, HD~35079 and HD~55606 are the strongest candidates for follow-up as
dynamically plausible triple-assisted interaction systems. HD~205329 remains
important but requires caution because of its astrometric interpretation.
HD~41335 is particularly valuable as a diagnostic system, because its period,
PMa, and speckle constraints appear to trace a more complex architecture than a
single simple triple.

A broader \textit{Gaia}-based comparison of the kinematics of Be stars and non-emission
B stars, including the incidence of runaway or walkaway candidates, would
provide an independent test of whether triple-assisted interaction leaves a
measurable population-level signature.

\section{Conclusions}
\label{sec:conclusions}

We have presented a dynamical plausibility study of Be-star multiple systems in
which speckle-resolved companions, literature inner periods, and
PMas are interpreted within a hierarchical-triple framework.
The analysis does not attempt to derive unique orbital solutions. Instead, it
identifies regions of synthetic parameter space that are simultaneously
compatible with the adopted dynamical filters and with the available
observational constraints. Our main conclusions are as follows.

The synthetic population confirms that a non-negligible subset of stable
hierarchical triples can operate in the secular regime with
$t_{\rm KL}<t_{\rm MS}$. Thus, outer companions in the subarcsecond-to-few-arcsecond range probed by speckle interferometry can be dynamically relevant
tertiaries rather than merely wide visual neighbours. 
Several systems in the sample can be represented by stable hierarchical
architectures that are both compatible with the available observational
constraints and capable of entering the KL-active regime.

The mass-aware refined simulations identify HD~35079 and HD~55606 as
the strongest dynamically plausible systems in the present sample. In both
cases, the observed inner period and the speckle-resolved projected separation
can be reproduced simultaneously by stable, secularly active hierarchical
configurations. The matched fractions are
$f_{\rm match}=8.79\times10^{-3}$ for HD~35079 and
$f_{\rm match}=4.08\times10^{-3}$ for HD~55606. Their median normalized
KL timescales remain below unity,
$t_{\rm KL}/t_{\rm MS}=0.0091$ and $0.0285$, respectively, indicating that
secular evolution can occur within the main-sequence lifetime of the Be
primary.

HD~205329 remains a dynamically plausible but cautionary case. The
PMa-plus-speckle constraints can be reproduced with
$f_{\rm match}=1.8\times10^{-3}$ and a median
$t_{\rm KL}/t_{\rm MS}=0.38$. Its interpretation depends more strongly on the
PMa-inferred orbital scale and therefore requires additional astrometric or
spectroscopic confirmation.

HD~41335 provides a useful diagnostic of constraint tension. The
period-plus-speckle and PMa-plus-speckle scenarios are viable separately, with
matched fractions of $9.3\times10^{-4}$ and $2.2\times10^{-3}$, respectively.
However, no accepted configurations are found when the period, PMa scale, and
speckle separation are imposed simultaneously under the adopted 30 per cent
tolerances. This suggests that the observed period and the PMa signal likely
trace different dynamical scales, or that the system requires a more complex
architecture than the simple hierarchical triple model explored here.

The inner-periastron interaction proxy is satisfied by the accepted
mass-aware scenarios at the order-of-magnitude level. The median values of
$r_{\rm peri,in}/(R_1+R_2)$ are close to or below unity for the viable refined
scenarios, supporting the possibility that secular eccentricity excitation can
bring the inner binary into an interaction regime.

The angular-momentum budget should be interpreted relative to the star
being spun up. In the adopted scaling, the median value
$\Delta M_{\rm req}/M\simeq4.8\%$ is normalized to the mass of the Be component.
Physically, $\Delta M_{\rm req}$ represents the mass that would need to be added to the Be star with enough specific angular momentum to spin it up towards near-critical rotation. This added mass can indeed be interpreted as mass donated by the inner companion, if the channel is mass transfer, but it does not necessarily equal all the mass lost by the donor. If the transfer is conservative and efficient, then $\Delta M_{\rm req}$ approximates the mass actually accreted by the Be star. However, if there is mass loss from the system, non-conservative accretion, material ejection, or a merger-type interaction, then $\Delta M_{\rm req}$ works as a mass scale with high angular momentum, rather than a literal prediction of how much the secondary star lost.

Finally, the global angular-momentum budget indicates that the spin-up
requirement is compatible with modest fractional mass accretion under plausible
assumptions. The median requirement is
$\Delta M_{\rm req}/M\simeq4.8\%$, with a 16--84 percentile range of
approximately $2.7$--$8.5\%$. When scaled to the accepted samples,
this corresponds to median accreted masses of
$\sim0.24\,M_\odot$ for HD~35079,
$\sim0.30\,M_\odot$ for HD~55606,
$\sim0.53\,M_\odot$ for HD~205329, and
$\sim0.74$--$0.79\,M_\odot$ for the viable HD~41335 partial-constraint
scenarios.

The results support triple-assisted interaction as a dynamically plausible
pathway towards Be-star spin-up. The next observational step is to obtain
additional orbital information: continued speckle monitoring to constrain the
outer companions, high-resolution spectroscopy to confirm or refine inner
periods, and improved astrometry to clarify the interpretation of PMa signals.
Such observations will be essential for moving from dynamical plausibility to a
fully constrained evolutionary history for the best candidate systems.

\begin{acknowledgements}
We are particularly grateful to the referee
for the exceptionally careful and constructive review
of our manuscript. Co-funded by the European Union (Project 101183150 - OCEANS). This work was partially supported by the PAPIIT Project IF100526: \textit{Computational Infrastructure for Astronomical Data Science at the Ensenada Campus of the Institute of Astronomy UNAM}. M.K. and O.V.M. acknowledge financial support from the Czech Science Foundation
(GA \v{C}R, grant number 25-17532S). The Astronomical Institute of the Czech Academy of Sciences is supported by the project RVO:67985815.
\end{acknowledgements}

\bibliographystyle{aa}
\bibliography{bibliography}

\begin{appendix}
\nolinenumbers

\section{Observational sample inventory}
\label{sec:ap-observational-sample}

Table~A.1 lists the 33 unique targets used in the observational
classification. Systems with multiple resolved pairs, such as
HD~161004 and HD~76838, contribute more than one separation
to the speckle distributions but are counted only once in the
target inventory.

\begin{table}[ht!]
\caption{Observed targets used in the dynamical experiment.}
\label{tab:observational_sample}
\centering
\begin{tabular}{lll}
\hline\hline
Target & Constraint group & Source \\
\hline
HD~17520  & $\rho$-$\mathrm{only}$ & G25 \\
HD~37115  & $\rho$-$\mathrm{only}$ & G25 \\
HD~45828  & $\rho$-$\mathrm{only}$ & G25 \\
HD~51893  & $\rho$-$\mathrm{only}$ & G25 \\
HD~72067  & $\rho$-$\mathrm{only}$ & S20 \\
HD~159032 & $\rho$-$\mathrm{only}$ & S20 \\
HD~161004 & $\rho$-$\mathrm{only}$ & S20 \\
HD~168135 & $\rho$-$\mathrm{only}$ & S20 \\
HD~174886 & $\rho$-$\mathrm{only}$ & S20 \\
HD~179343 & $\rho$-$\mathrm{only}$ & S20 \\
HD~219523 & $\rho$-$\mathrm{only}$ & G25 \\
HD~232940 & $\rho$-$\mathrm{only}$ & G25 \\
\hline
HD~18231  & PMa-only & K19 \\
HD~23478  & PMa-only & K19 \\
HD~29096  & PMa-only & K19 \\
HD~32811  & PMa-only & K19 \\
HD~36012  & PMa-only & K19 \\
HD~37266  & PMa-only & K19 \\
HD~39018  & PMa-only & K19 \\
HD~54551  & PMa-only & K19 \\
HD~57539  & PMa-only & K19 \\
HD~180126 & PMa-only & K19 \\
HD~183035 & PMa-only & K19 \\
HD~204860 & PMa-only & K19 \\
HD~215837 & PMa-only & K19 \\
\hline
HD~35079 & $P+\rho$ & Lit.; G25 \\
HD~55606 & $P+\rho$ & Lit.; G25 \\
\hline
HD~37330  & $\mathrm{PMa}+\rho$ & K19; G25 \\
HD~41600  & $\mathrm{PMa}+\rho$ & K19; G25 \\
HD~71510  & $\mathrm{PMa}+\rho$ & K19; S20 \\
HD~205329 & $\mathrm{PMa}+\rho$ & K19; G25 \\
\hline
HD~41335 & $P+\mathrm{PMa}+\rho$ & Lit.; K19; G25 \\
HD~76838 & $P+\mathrm{PMa}+\rho$ & Lit.; K19; S20 \\
\hline
\end{tabular}
\tablefoot{
The table lists the 33 unique targets used in the observational classification
of Table~\ref{tab:full_target_sample}. The counts refer to targets rather than
individual visual pairs or speckle detections. Systems with multiple resolved
visual pairs, such as HD~161004 or HD~76838, can contribute more than one
speckle separation to the separation-distribution comparisons, but they are
counted only once in the target inventory. Source codes:
S20 = \citet{2020AJ....159..132S};
G25 = \citet{2025AJ....169..251G};
K19 = \citet{2019AA...623A..72K};
Lit. = additional spectroscopic or binary-orbit literature cited in the main text.
}
\end{table}

\FloatBarrier

\section{Speckle detectability window}
\label{sec:ap-speckle-detectability}

\begin{figure}[!t]
    \centering
    \includegraphics[width=\columnwidth]{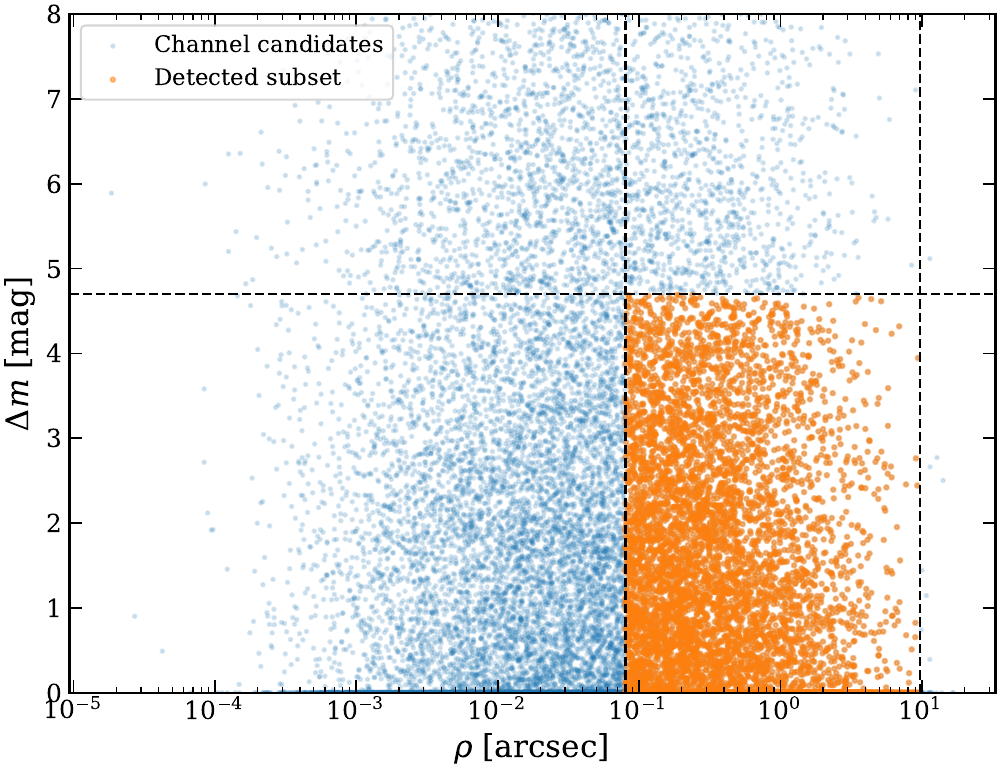}
    \caption{
    Adopted angular-separation and magnitude-contrast limits for speckle detection. Synthetic outer companions are required to fall within this
    observationally accessible region before they are compared with the
    target-specific speckle separations. The window summarizes the practical
    separation and contrast range of the speckle observations used in this
    analysis. The vertical dashed lines mark the angular-separation interval,
$0.08''\leq\rho\leq10''$, and the horizontal dashed line marks the adopted
contrast threshold, $\Delta m=4.7$ mag.}
    \label{fig:speckle_detectability_window}
\end{figure}

Figure~\ref{fig:speckle_detectability_window} illustrates the adopted
speckle-detectability window in angular separation and magnitude difference.
The vertical dashed lines mark the lower and upper angular-separation limits
used in the simulations, while the horizontal dashed line marks the adopted
contrast limit. Together, these boundaries define the region in which a
synthetic outer companion is considered accessible to the speckle surveys used
in this work. The window provides an observationally motivated approximation
to the separations and magnitude differences over which companions can be
compared consistently with the Be-star speckle detections.

\FloatBarrier
\section{Hierarchy ratio and secular timescale}
\label{sec:ap-hierarchy-timescale}

\begin{figure}[!t]
    \centering
    \includegraphics[width=\columnwidth]{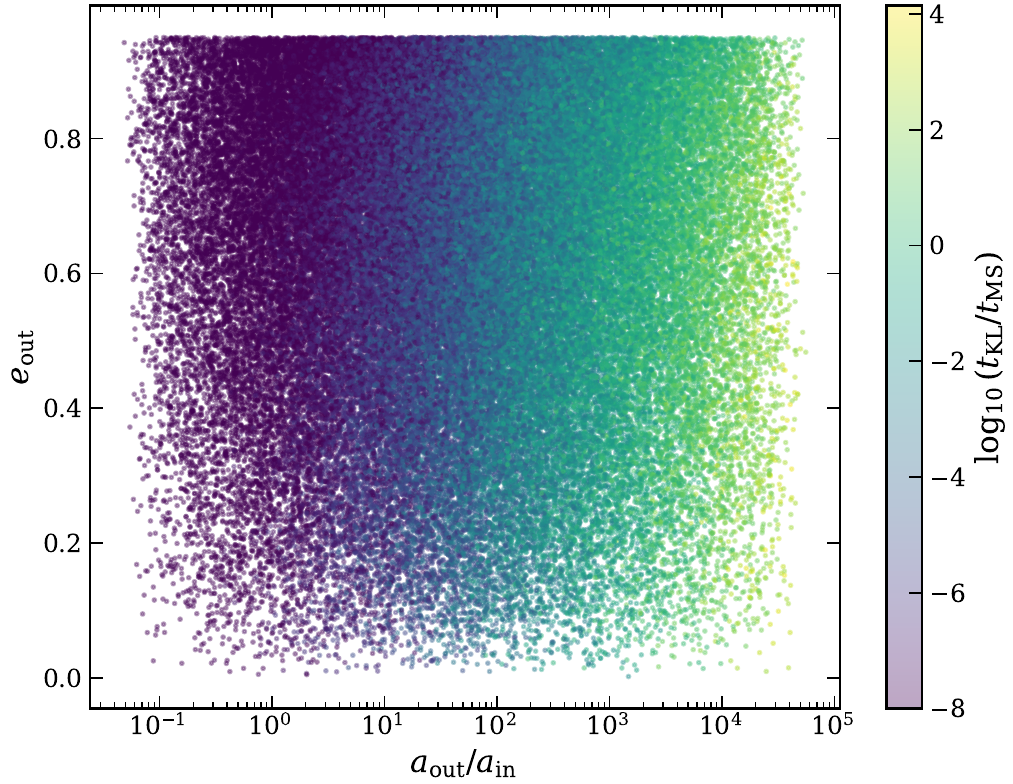}
    \caption{
    Hierarchy ratio and secular-timescale diagnostic for the synthetic
    hierarchical triples. The horizontal axis shows $a_{\rm out}/a_{\rm in}$,
    the vertical axis shows the outer eccentricity ($e_{\rm out}$), and the colour
    scale gives $\log_{10}(t_{\rm KL}/t_{\rm MS})$. Negative values correspond
    to systems for which the KL timescale is shorter than the
    main-sequence lifetime of the Be component. The figure illustrates how the
    secular timescale varies across the synthetic architecture space before
    the target-specific observational constraints are imposed.
    }
    \label{fig:stability_timescale_map}
\end{figure}

Figure~\ref{fig:stability_timescale_map} provides a complementary diagnostic of
the synthetic hierarchical architectures. The horizontal axis shows the
hierarchy ratio, $a_{\rm out}/a_{\rm in}$, while the vertical axis shows the
outer eccentricity, $e_{\rm out}$. The colour scale gives
$\log_{10}(t_{\rm KL}/t_{\rm MS})$, where $t_{\rm KL}$ is the KL
timescale and $t_{\rm MS}$ is the adopted main-sequence lifetime of the Be
primary.

This representation is useful because the secular timescale depends strongly
on the separation hierarchy. Wider outer orbits generally produce longer
secular timescales, whereas more compact hierarchical configurations can drive
faster secular evolution. The outer eccentricity also contributes through the
factor $(1-e_{\rm out}^{2})^{3/2}$ in the adopted KL-timescale prescription, so
more eccentric outer orbits can shorten $t_{\rm KL}$ for a fixed hierarchy
ratio.

The figure should be read as a diagnostic map rather than as an additional
selection criterion. Configurations with
$\log_{10}(t_{\rm KL}/t_{\rm MS})<0$ have KL timescales shorter than the
main-sequence lifetime of the Be star. Therefore, this diagram helps
visualise the region of synthetic parameter space in which the secular channel
can operate efficiently before the target-specific period, PMa, and
speckle-separation constraints are applied.

\FloatBarrier
\section{Projected-separation distributions}
\label{sec:ap-rho-distributions}

\begin{figure*}[!t]
    \centering
    \includegraphics[width=0.71\textwidth]{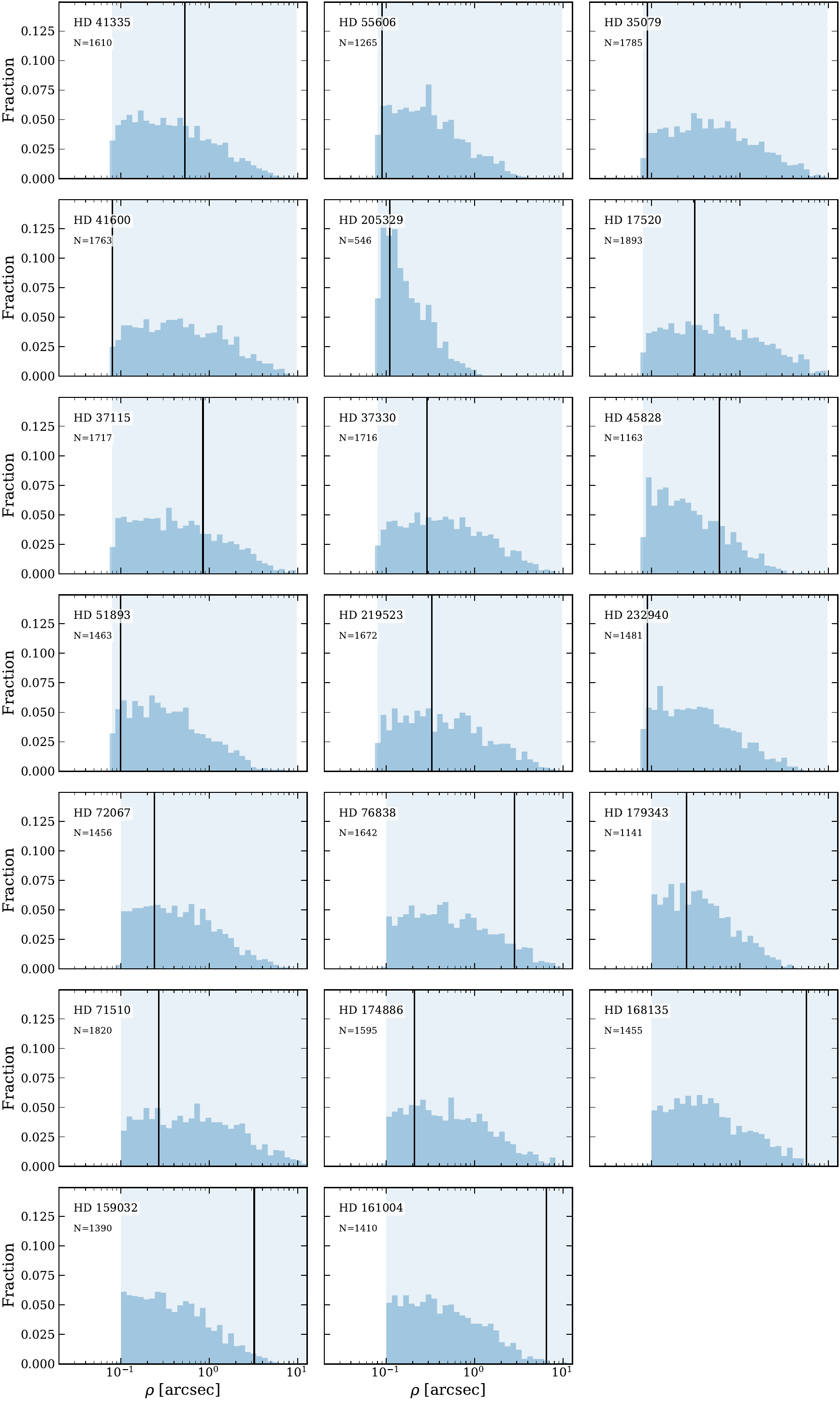}
    \caption{
    Full projected-separation distributions for the accepted synthetic systems.
    Each panel shows the distribution of $\rho_{\rm model}$ for one target after the synthetic projected physical separations are converted to angular separations
using the adopted distance of that system, and after the dynamical and
object-specific observational filters are applied. The value
$N$ indicated in each panel is the number of accepted synthetic configurations
used to build that distribution. The vertical black
line marks the observed speckle separation, $\rho_{\rm obs}$, and
    the shaded region indicates the adopted speckle-detection window. A
    compact summary of these distributions is shown in
    Fig.~\ref{fig:rho_summary}.
    }
    \label{fig:rho_panels_appendix}
\end{figure*}

Figure~\ref{fig:rho_panels_appendix} shows the full projected-separation
distributions for the accepted synthetic systems. These distributions were
constructed after applying the dynamical filters and the object-specific
observational constraints described in Sects.~\ref{sec:observational_constraints}
and~\ref{sec:dynamical_framework}. The observed speckle separations used as
reference values were taken from our Be-star speckle surveys
\citet{2020AJ....159..132S} and \citet{2025AJ....169..251G}. The vertical line
in each panel marks the corresponding observed speckle separation, while the
shaded band indicates the adopted detection window.

\end{appendix}
\end{document}